\documentclass[letterpaper,journal]{IEEEtran}

\usepackage[thmmarks]{ntheorem} 
\usepackage{amsmath,amsfonts}
\usepackage{algorithmic}
\usepackage{algorithm}
\usepackage{array}
\usepackage{textcomp}
\usepackage{todonotes}
\usepackage{stfloats}
\usepackage{url}
\usepackage{verbatim}
\usepackage{graphicx}
\usepackage{epsfig}
\usepackage{cite}
\usepackage{pifont}
\usepackage{amsmath}
\usepackage{booktabs}
\usepackage{subcaption}
\usepackage{tabularx}
\usepackage{makecell}
\usepackage{subfloat}
\usepackage{bm}
\usepackage{xcolor}
\usepackage{nccmath}
\usepackage[titles,subfigure]{tocloft}
\makeatletter
\renewcommand{\maketag@@@}[1]{\hbox{\m@th\normalsize\normalfont#1}}%
\makeatother
\allowdisplaybreaks
\begin{document}
	
	%\title{
		%\textcolor{black}{Enhancing VSG Transient Synchronization Stability Under Grid Frequency Disturbances}}
	
	\title{
		\textcolor{black}{Enhancing Transient Stability of VSG-Controlled Converters Under Grid Frequency Disturbances}}

	\author{Bingfang Li,~\IEEEmembership{Member,~IEEE}, Yubo Zhang,~\IEEEmembership{Member,~IEEE}, Xiaotong Wang,\\   Songhao Yang,~\IEEEmembership{Senior Member,~IEEE}, Zhiguo Hao,~\IEEEmembership{Senior Member,~IEEE}, Federico Milano,~\IEEEmembership{Fellow,~IEEE}%
		% <-this % stops a space
	\thanks{\textit{Corresponding author: Zhiguo Hao.}}%
	\thanks{B. Li, Y. Zhang, X. Wang, S. Yang, and Z. Hao are
		with the School of Electrical Engineering, Xi'an Jiaotong
		University, Xi'an, China (e-mail: zhghao@xjtu.edu.cn).}%
	\thanks{F. Milano is with the School of Electrical and
		Electronic Engineering, University College Dublin,
		Dublin 4, Ireland (e-mail: federico.milano@ieee.org).}%
	}
	
		%\thanks{This work was supported by the Science and Technology Project of State Grid Jibei Electric Power Co.(52018K22001D).}% <-this % stops a space
		%\thanks{B. Li, S. Yang, Z. Hao and B. Zhang are with Xi’an Jiaotong University, Xi'an, China (e-mail: libingfang@stu.xjtu.edu.cn, {songhaoyang, zhghao, bhzhang}@xjtu.edu.cn,). H. Xie and T. Zhao are with State Grid Jibei Electric Power Co.}
	% The paper headers
	%\markboth{Journal of \LaTeX\ Class Files,~Vol.~14, No.~8, August~2021}%
	%{Shell \MakeLowercase{\textit{et al.}}: A Sample Article Using IEEEtran.cls for IEEE Journals}
	%\IEEEpubid{0000--0000/00\$00.00~\copyright~2021 IEEE}
	% Remember, if you use this you must call \IEEEpubidadjcol in the second
	% column for its text to clear the IEEEpubid mark.
	\maketitle
	\begin{abstract}
		Strong damping feedback can substantially improve the transient synchronization stability of renewable energy converters employing virtual synchronous generator (VSG) control. However, conventional damping feedback uses the nominal frequency as its reference. When the frequency of a weak grid deviates from nominal,
		strong damping tends to keep the VSG frequency near nominal.
		This can sustain a frequency difference between the VSG and
		the grid and compromise transient synchronization. To address this problem, this paper proposes a complex-frequency-based method to estimate the grid reference frequency from local terminal measurements. The estimated frequency replaces the nominal reference in the VSG synchronization loop, so that damping acts on the frequency difference between the VSG and the real grid.  Controller hardware-in-the-loop tests validate the effectiveness of the proposed method under grid frequency disturbances and its robustness to grid parameter uncertainty and adverse grid conditions.
	\end{abstract}

	\begin{IEEEkeywords}
	\textcolor{black}{Complex frequency, grid frequency disturbances, grid frequency estimation, transient synchronization stability, virtual synchronous generator}
	\end{IEEEkeywords}
	
	\section{Introduction}
	\IEEEPARstart{T}{he}
		large-scale integration of renewable generation through grid-following (GFL) converters has reduced system inertia and short-circuit capacity, making modern power systems more vulnerable to wideband oscillations and voltage or frequency violations~\cite{Strunz2023}. Against this backdrop, grid-forming (GFM) control has attracted increasing attention for renewable energy converters. GFM converters establish an autonomous voltage source at the grid connection point with self-regulated voltage and frequency~\cite{Matevosyan2023}. Among various GFM strategies, virtual synchronous generator (VSG) control emulates synchronous generator (SG) dynamics, enabling renewable energy converters to provide virtual inertia, damping, and voltage support~\cite{Leon2024,Matevosyan2023}.
		
		Like synchronous generators, VSG-controlled converters may lose synchronism following large disturbances~\cite{Lei2023}. Considerable efforts have therefore been devoted to improving their transient synchronization stability.
	Some studies reshape the post-fault power angle relationship by adjusting the converter voltage, active power response, or virtual impedance, thereby modifying the acceleration and deceleration areas and enlarging the operating region with a stable post-fault equilibrium~\cite{Pan2020,Zhang2023CCPAC,Wang2025VI}. 
Other studies improve transient synchronization stability by adaptively tuning control parameters, such as power feedback gains, virtual inertia, damping coefficients, and current-saturation angles, thereby reshaping the transient power imbalance and energy dissipation~\cite{Lei2023,Xiong2021,Luo2025}.
	For multi-VSG systems, controller limit design based on 
	transient energy functions has also been developed to prevent 
	excessive frequency excursions and promote transient energy 
	dissipation~\cite{Li2025}. These studies have provided valuable 
	guidance for designing VSG controls to improve transient stability.
	%However, most existing studies assume an infinite bus at the 
	%nominal frequency as the external grid. This assumption becomes 
	%overly restrictive in weak power systems, where severe 
	%disturbances can cause substantial deviations of the grid 
	%frequency from its nominal value.
	
	Overall, although VSGs inherit instability mechanisms similar to those of SGs through swing equation emulation, their flexible control allows substantial improvements in transient stability. 
	Among these methods, frequency deviation feedback is one of the simplest
	and most widely used approaches, providing fast primary frequency
	response or strong damping \cite{Xiong2021,Lei2023,Li2023Global}.
 However, most of these studies represent the external grid as an infinite bus operating at the nominal frequency. This assumption is restrictive for weak power systems, where severe disturbances can cause substantial grid frequency deviations.

	 Following a sudden load increase or generation loss, the resulting power imbalance can drive the grid frequency away from nominal \cite{Milano2018,Zhang2024}. Primary frequency control arrests the frequency decline, while secondary control restores the nominal frequency over a longer timescale \cite{NERC2021}. The grid frequency may therefore remain off-nominal throughout the transient and early recovery stages~\cite{Hao2026}.
		
		An off-nominal grid frequency does not in itself prevent synchronization in conventional power systems. SGs synchronize through electromagnetic interactions with the connected grid and may regain synchronism at an off-nominal frequency \cite{Kundur1994,Demello1969}. Frequency control subsequently restores the interconnected system to the nominal frequency.
	%However, GFM control implements frequency regulation or damping in software, typically using the nominal frequency as its reference \cite{Matevosyan2023,Lei2023}. Its fast response can drive the converter frequency toward nominal while the grid frequency remains displaced. Strong VSG frequency feedback may therefore sustain a frequency difference between the converter and the grid, causing the power angle to increase and threatening transient synchronization stability.
	However, VSG-controlled converters commonly use the nominal frequency as the reference for damping or frequency regulation\cite{Matevosyan2023,Lei2023}. When the grid frequency deviates from nominal, strong feedback toward this fixed reference can sustain a frequency difference between the converter and the grid, causing the power-angle deviation to accumulate and potentially leading to loss of synchronism.

	Accurately estimating the actual grid frequency and using it as the VSG synchronization reference is key to addressing this problem. 
    Wide-area monitoring systems (WAMS) can provide remote grid frequency measurements, but require communication infrastructure and introduce communication delays \cite{DiezMaroto2019}. Local phase-locked loop (PLL) measurements offer an alternative, but face two limitations. First, frequency divider theory  \cite{Milano2017} indicates that the terminal frequency can be dominated by the local VSG frequency in weak grids, so that it may not accurately represent the remote grid frequency. Second, when PLL measurements are used for feedback control, inappropriate PLL tuning can compromise system stability \cite{Zhou2022PLL}. Further challenges arise under practical operating conditions. VSGs may switch between constant voltage control (CVC) and current limiting control (CLC) during transients, while GFM converters often operate alongside GFL-controlled renewable energy converters at the same point of common coupling \cite{Li2026}. Estimating the grid frequency from local measurements while accounting for these operating conditions therefore remains an important challenge. 
	
	This paper proposes a method based on complex frequency theory to enhance VSG synchronization stability by adapting its frequency reference to the actual grid frequency. The main contributions are as follows.
	
	1) Transient energy analysis reveals how time-varying grid frequency can introduce effective negative damping into the relative dynamics between the VSG and the grid.
	
	2) A method based on complex frequency theory is developed to estimate the equivalent Thevenin grid frequency without remote communication. It accommodates transitions between normal and current-limiting control of GFM converters, as well as parallel operation with GFL converters.
	
	3) The estimated grid frequency is used as an adaptive reference for VSG frequency feedback. This allows strong damping to improve stability during grid frequency dips while retaining its benefits during voltage sags. The proposed method is also robust to grid parameter uncertainty and variations in filter settings.

\section{Impact of Grid Frequency Deviations on VSG Transient Synchronization Stability}

%Classical VSG transient-stability analysis usually assumes an infinite-bus grid whose frequency is fixed at the nominal value $\omega_N$. 
%In weak grids, however, severe disturbances can drive the grid frequency away from $\omega_N$ during the VSG synchronization process. As a result, $\omega_\mathrm{vsg}-\omega_N$ no longer represents the true VSG-grid relative frequency $\omega_\mathrm{vsg}-\omega_g$.
%Because conventional VSG damping feedback is referenced to $\omega_N$ rather than to the actual grid frequency, grid frequency drift introduces an additional disturbance into the VSG synchronization dynamics. This section analyzes how this disturbance affects VSG transient stability and derives the corresponding stability boundary.

\subsection{System Overview}
\begin{figure}[!t]
	\centering
	\includegraphics[width=\columnwidth,trim=3mm 0mm 0mm 0mm,clip]{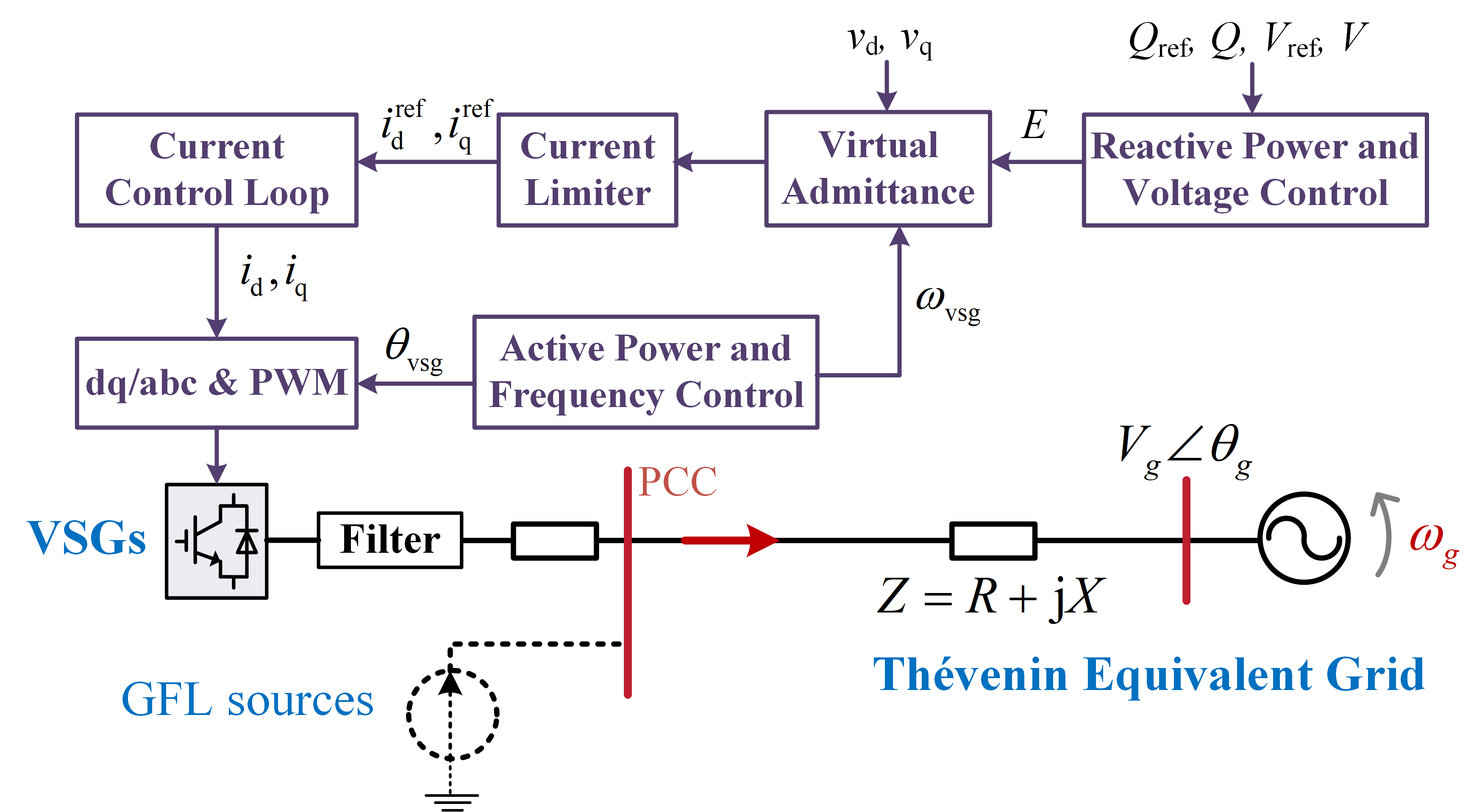}
	\caption{Topology of the grid-connected VSG system.}
	\label{TOPOLOGY}
\end{figure}

Consider the grid-connected system shown in Fig.~\ref{TOPOLOGY}.
The VSG has an internal voltage $\bar{E}=Ee^{j\theta_e}$
and a virtual frequency
$\omega_{\mathrm{vsg}}=\dot{\theta}_e$.
The outputs of the VSG and parallel GFL sources
are combined at the point of common coupling (PCC).
The connection from the PCC to the Th\'evenin equivalent grid
includes step-up transformers and transmission lines,
collectively modeled as the equivalent impedance
$\bar{Z}=R+jX$.
The positive-sequence PCC voltage and total current flowing
from the PCC toward the grid are denoted by
$\bar{V}=Ve^{j\theta_v}$ and $\bar{I}=Ie^{j\theta_i}$,
respectively.
The Th\'evenin equivalent grid voltage is
$\bar{V}_g=V_ge^{j\theta_g}$, with instantaneous frequency
$\omega_g=\dot{\theta}_g$.
Following a large disturbance, $\omega_g$ may deviate
substantially from the nominal frequency $\omega_N$.

\subsection{Mechanism of Synchronization Instability Under Grid Frequency Decline}
To explain VSG transient synchronization stability during grid frequency
dips, the theoretical analysis uses a reduced-order model.
The controller hardware-in-the-loop (CHIL) tests, however, use the full nonlinear switching model
to validate the proposed method.

Let $\delta=\theta_e-\theta_g$ denote the VSG power angle
relative to the grid, with relative frequency
$\Delta\omega=\dot{\delta}=\omega_{\mathrm{vsg}}-\omega_g$.
The frequency deviation used for feedback is
$e_\omega=\omega_{\mathrm{vsg}}-\omega_{\mathrm{ref}}$,
where $\omega_{\mathrm{ref}}$ is the feedback reference.

Frequency feedback can be implemented through the
active power reference, the damping loop, or both.
When these paths use the same frequency reference,
their proportional and integral contributions can be combined.
The dynamic equations of the VSG synchronization loop are
\begin{align}
	M\dot e_\omega
	&=P_m-P_e(\delta)-D e_\omega-K_i z,
	\label{eq:vsg_swing}\\
	\dot z&=e_\omega,
	\label{eq:frequency_integrator}\\
	\omega_{\mathrm{vsg}}&=e_\omega+\omega_{\mathrm{ref}},
	\label{eq:vsg_frequency_reconstruction}
\end{align}
where $M$ is the virtual inertia, $P_m$ is the active
power reference before frequency-feedback correction,
and $P_e(\delta)$ is the electrical output power.
The coefficients $D$ and $K_i$ represent the combined
proportional and integral gains, respectively.
Conventional control uses $\omega_{\mathrm{ref}}=\omega_N$,
so that
$e_\omega=\Delta\omega+(\omega_g-\omega_N)$.
Since $\omega_N$ is constant,
$\dot e_\omega=\Delta\dot{\omega}+\dot{\omega}_g$,
which gives
\begin{align}
	M\Delta\dot{\omega}
	&=P_m-P_e(\delta)-D\Delta\omega-K_i z
	\notag\\
	&\quad-D(\omega_g-\omega_N)-M\dot{\omega}_g,
	\label{eq:relative_swing}\\
	\dot{\delta}&=\Delta\omega,
	\label{eq:angle_eq}\\
	\dot z&=\Delta\omega+(\omega_g-\omega_N).
	\label{eq:relative_integrator}
\end{align}

Let \(\delta_s\) denote the stable equilibrium point angle.
The transient energy of the relative motion is defined as
\begin{equation}
	W(\delta,\Delta\omega)
	=\frac{1}{2}M\Delta\omega^2
	+\int_{\delta_s}^{\delta}
	\left[P_e(\xi)-P_m\right]\,d\xi.
	\label{eq:energy}
\end{equation}

Differentiating \eqref{eq:energy} and substituting \eqref{eq:relative_swing} and \eqref{eq:angle_eq} gives
\begin{align}
	\dot W
	&=\Delta\omega
	\left[M\Delta\dot{\omega}+P_e(\delta)-P_m\right]
	\notag\\
	&=-D\Delta\omega^2-K_i z\Delta\omega
	\notag\\
	&\quad-D(\omega_g-\omega_N)\Delta\omega
	-M\dot{\omega}_g\Delta\omega.
	\label{eq:Vdot}
\end{align}
The proportional feedback produces both the dissipative term \(-D\Delta\omega^2\) and the reference mismatch term \(-D(\omega_g-\omega_N)\Delta\omega\). The reference mismatch also enters the integral feedback through \(\dot z=\Delta\omega+(\omega_g-\omega_N)\).

During a grid frequency decline below nominal,
$-D(\omega_g-\omega_N)\Delta\omega$ and
$-M\dot{\omega}_g\Delta\omega$ both inject energy
when $\Delta\omega>0$, opposing the damping dissipation. As the VSG approaches $\omega_g$ during the frequency dip,
$z$ decreases once $\omega_{\mathrm{vsg}}<\omega_N$.
If $z$ becomes negative while $\Delta\omega>0$,
$-K_i z\Delta\omega$ also injects energy.
Together, these energy inputs can exceed damping
dissipation, causing transient energy to grow and
threatening synchronization stability.

If $\omega_{\mathrm{ref}}=\omega_g$, then $e_\omega=\Delta\omega$.
Damping therefore acts on the actual frequency difference
between the VSG and the grid.
This choice can improve transient synchronization stability
during a grid frequency dip.

\section{Grid Frequency Estimation and Adaptive Frequency Feedback}
\label{sec:mode7_estimation}

This section develops a local estimator of the Th\'evenin grid
frequency based on complex frequency theory.
The estimated frequency then replaces the nominal reference
in VSG frequency reconstruction and feedback.

\subsection{Independent Reference Frame and Network Model}
To avoid introducing PLL tracking errors into the estimator,
the PCC voltage and current are transformed into an independent
$dq$ reference frame with angle
\begin{equation}
	\theta_0(t)=\theta_0(t_0)+\omega_N(t-t_0),
	\label{eq:m7_frame}
\end{equation}
where $\theta_0(t)$ is the angle of the independent $dq$
reference frame at time $t$ and $\omega_N$ is the nominal
angular frequency. The initial time and angle are denoted
by $t_0$ and $\theta_0(t_0)$, respectively.

The positive-sequence PCC voltage and current in this frame
are expressed as
\begin{align}
	\bar V^{dq}&=\bar V e^{-j\theta_0}=v_d+jv_q,\notag\\
	\bar I^{dq}&=\bar I e^{-j\theta_0}=\imath_d+j\imath_q,
	\label{eq:m7_dq_vectors}
\end{align}
where $v_d$ and $v_q$ are the $d$- and $q$-axis voltage
components, and $\imath_d$ and $\imath_q$ are the corresponding
current components.

The Th\'evenin grid voltage in the same frame is
\begin{equation}
	\bar V_g^{dq}
	=\bar V_g e^{-j\theta_0}
	=V_g e^{j(\theta_g-\theta_0)},
	\label{eq:m7_grid_voltage_dq}
\end{equation}
where $V_g$ and $\theta_g$ are the magnitude and phase angle
of the Th\'evenin grid voltage, respectively.

Retaining the inductor dynamics gives
\begin{equation}
	\bar V-\bar V_g=R\bar I+L\dot{\bar I},
	\label{eq:m7_stationary_network}
\end{equation}
where $R$ is the equivalent resistance and $L=X/\omega_N$
is the equivalent inductance, with $X$ denoting the reactance
at the nominal frequency.

Substituting
$\dot{\bar I}=e^{j\theta_0}
(\dot{\bar I}^{dq}+j\omega_N\bar I^{dq})$
into \eqref{eq:m7_stationary_network} gives
\begin{equation}
	\bar V_g^{dq}
	=\bar V^{dq}-(R+jX)\bar I^{dq}
	-L\dot{\bar I}^{dq}.
	\label{eq:m7_dq_network}
\end{equation}

\subsection{Grid Frequency Estimation Based on Complex Frequency}

The complex frequency of the Th\'evenin
grid voltage in the stationary reference frame is defined as~\cite{Milano2022}
\begin{equation}
	\bar\eta_g
	=\frac{\dot{\bar V}_g}{\bar V_g}
	=\rho_g+j\omega_g,
	\label{eq:vg_complex_frequency}
\end{equation}
where the real part $\rho_g=\dot V_g/V_g$ describes the relative rate
of change of the voltage magnitude, and the imaginary part
$\omega_g=\dot\theta_g$ is the grid angular frequency.

Combining \eqref{eq:m7_grid_voltage_dq} and
\eqref{eq:vg_complex_frequency} gives the complex frequency
relative to the independent $dq$ reference frame as
\begin{equation}
	\frac{\dot{\bar V}_g^{dq}}{\bar V_g^{dq}}
	=\bar{\eta}_g-j\omega_N
	=\rho_g+j(\omega_g-\omega_N).
	\label{eq:vg_complex_frequency_dq}
\end{equation}
Taking the imaginary part of \eqref{eq:vg_complex_frequency_dq}
gives the grid angular frequency as
\begin{equation}
	\omega_g
	=\omega_N+\operatorname{Im}
	\left(
	\frac{\dot{\bar V}_g^{dq}}{\bar V_g^{dq}}
	\right).
	\label{eq:vg_frequency_imag}
\end{equation}

From \eqref{eq:m7_dq_network}, the $d$- and $q$-axis
components of the reconstructed grid voltage are
\begin{align}
	v_{gd}&=v_d-R\imath_d+X\imath_q-L\dot{\imath}_d,\notag\\
	v_{gq}&=v_q-R\imath_q-X\imath_d-L\dot{\imath}_q.
	\label{eq:vg_reconstructed_components}
\end{align}

Substituting $\bar V_g^{dq}=v_{gd}+jv_{gq}$ into
\eqref{eq:vg_frequency_imag} gives the grid angular
frequency estimate as
\begin{equation}
	\hat{\omega}_g
	=\omega_N+
	\frac{v_{gd}\dot v_{gq}-v_{gq}\dot v_{gd}}
	{v_{gd}^{2}+v_{gq}^{2}}.
	\label{eq:vg_frequency_estimator}
\end{equation}
Since the derivation does not depend on the VSG control law,
the same estimation formula applies to both CVC and CLC.

In principle, the reconstructed Th\'evenin voltage
$\bar V_g^{dq}$, rather than the local PCC voltage,
could also be supplied to a PLL or a frequency-locked loop (FLL)
to obtain $\hat{\omega}_g$.
However, a PLL or FLL is itself a closed-loop tracker
whose convergence depends on the voltage magnitude
and on its own gain tuning.
When the VSG switches to CLC, the reconstructed voltage
may change abruptly, and the tracking dynamics would be
coupled into the VSG synchronization loop through the
frequency feedback, introducing an additional stability
concern.
The proposed estimator avoids this issue because it is an
algebraic computation followed by low-pass filtering,
so its dynamics are determined only by the filter
bandwidth, which is examined in Section~\ref{sec:chil}.

\subsection{Adaptive Frequency Feedback and Implementation}

% Required packages:
% \usepackage{booktabs}
% \usepackage{tabularx}
% \usepackage{array}

\begin{table*}[!t]
	\centering
	\caption{Operating Conditions of the CHIL Tests}
	\label{tab:chil_cases}
	\footnotesize
	\renewcommand{\arraystretch}{1.15}
	\setlength{\tabcolsep}{5pt}
	
	\begin{tabularx}{\textwidth}{
			@{}l
			>{\raggedright\arraybackslash}X
			>{\raggedright\arraybackslash}p{0.30\textwidth}@{}
		}
		\toprule
		Case & Operating conditions & Disturbance \\
		\midrule
		
		C1
		& $P_{\mathrm{VSG}}=5~\mathrm{kW}$,
		$P_{\mathrm{GFL}}=4.67~\mathrm{kW}$.
		\newline
		CVC operation ($I_{\max}=3I_N$).
		\newline
		$R/X\approx0.0016$,
		$\mathrm{SCR}=2.39$,
		$f_{\omega}=5~\mathrm{Hz}$.
		& Frequency dip from $50$ to $48.3~\mathrm{Hz}$.
		\newline
		Maximum $|\mathrm{d}f_g/\mathrm{d}t|
		\approx1.8~\mathrm{Hz/s}$. \\
		
		C2
		& Power settings as in C1.
		\newline
		Transitions between CVC and CLC
		($I_{\max}=1.5I_N$).
		\newline
		$R/X\approx0.0016$,
		$\mathrm{SCR}=2.39$,
		$f_{\omega}=5~\mathrm{Hz}$.
		& Frequency dip as in C1. \\
		
		C3
		& As in C2, with $R/X=0.6$ and updated
		impedance parameters in the estimator.
		& Frequency dip as in C1. \\
		
		C4
		& As in C2, with $R/X=0.6$ and the original
		impedance parameters retained in the estimator.
		& Frequency dip as in C1. \\
		
		C5
		& As in C2, with $\mathrm{SCR}=1.5$,
		unchanged $R/X$, and updated impedance
		parameters in the estimator.
		& Frequency dip as in C1. \\
		
		C6
		& As in C2, with $\mathrm{SCR}=1.5$,
		unchanged $R/X$, and the original impedance
		parameters retained in the estimator.
		& Frequency dip as in C1. \\
		
		C7
		& As in C2, with $f_{\omega}=20~\mathrm{Hz}$.
		& Frequency dip as in C1. \\
		
		C8
		& As in C2, with $f_{\omega}=0.5~\mathrm{Hz}$.
		& Frequency dip as in C1. \\
		
		C9
		& As in C2, with $P_{\mathrm{VSG}}=8~\mathrm{kW}$.
		& Frequency dip as in C1 and balanced voltage sag
		($V_a=V_b=V_c=0.1~\mathrm{p.u.}$, hold $2~\mathrm{s}$).
	 \\
		
		C10
		& As in C2, with $P_{\mathrm{GFL}}=7.47~\mathrm{kW}$.
		& Frequency dip as in C1 and unbalanced voltage sag
		($V_a=0.30$, $V_b=0.285$,
		$V_c=0.315~\mathrm{p.u.}$, hold $2~\mathrm{s}$).
	\\
		
		C11
		& As in C2, with additional frequency PI feedback:
		$k_{p\omega}=50$ and $k_{i\omega}=500$.
		& Frequency dip as in C1. \\
		
		C12
		& Two VSGs with
		$P_1=5~\mathrm{kW}$ and $P_2=3~\mathrm{kW}$.
		No GFL converter.
		\newline
		$M_1=M_0$, $M_2=1.25M_0$,
		$D_1=4000$, $D_2=3000$,
		$I_{\max,1}=1.2I_N$,
		$I_{\max,2}=1.5I_N$.
		\newline
		Other grid and filter settings follow C2.
		& Frequency dip as in C1. \\
		
		\bottomrule
	\end{tabularx}

\end{table*}

\begin{figure}[!t]
	\centering
	\includegraphics[width=2.6in,trim=0mm 0mm 0mm 0mm,clip]{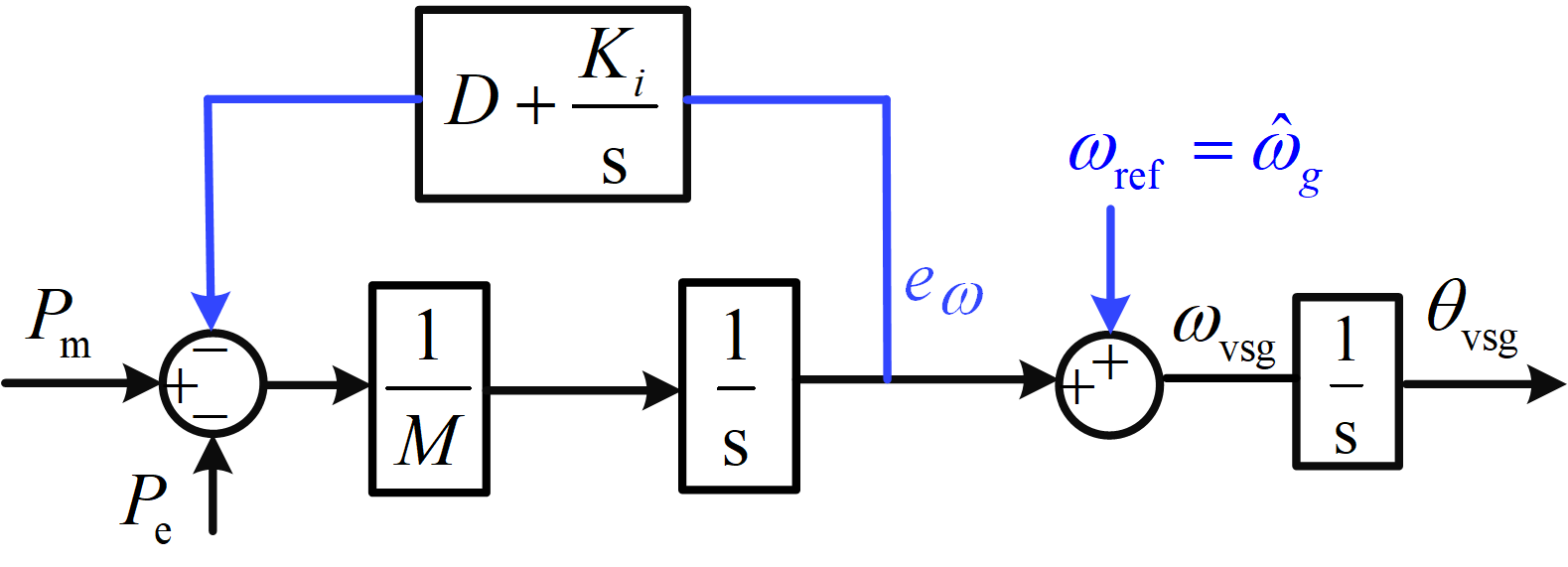}
	\caption{VSG active power control with adaptive frequency feedback.}
	\label{Control}
\end{figure}

As shown in Fig.~\ref{Control}, the estimated grid frequency
is used as the reference for frequency reconstruction
and feedback, giving
\begin{equation}
	\omega_{\mathrm{ref}}=\hat{\omega}_g,
	\qquad
	e_\omega=\omega_{\mathrm{vsg}}-\hat{\omega}_g.
	\label{eq:adaptive_frequency_reference}
\end{equation}
The feedback error is therefore the difference between the
VSG frequency and the estimated grid frequency.
The first integrator produces $e_\omega$ according to
\eqref{eq:vsg_swing}.
The controller then adds $\hat{\omega}_g$ to reconstruct
$\omega_{\mathrm{vsg}}$ through
\eqref{eq:vsg_frequency_reconstruction}.
The reconstructed absolute frequency drives the angle integrator.

The independent reference angle avoids introducing PLL tracking
dynamics or VSG angle dynamics into the coordinate transformation.
In a synchronous steady state, the reconstructed grid voltage
rotates at $\omega_g-\omega_N$ in this frame.
Near nominal frequency, its components vary slowly, which
facilitates filtering and numerical differentiation.
In the digital implementation, the required derivatives are
computed using backward differences followed by first-order
low-pass filtering.
The reconstructed voltage is filtered before differentiation,
and the resulting frequency estimate is further smoothed
before feedback.

Under ideal estimation, $\hat{\omega}_g=\omega_g$
and $e_\omega=\Delta\omega$.
Substituting this relation into \eqref{eq:vsg_swing} gives
\begin{equation}
	M\Delta\dot{\omega}
	=P_m-P_e(\delta)-D\Delta\omega-K_i z,
	\label{eq:adaptive_relative_swing}
\end{equation}
with $\dot z=\Delta\omega$.
Both the reference mismatch and the explicit grid frequency
derivative term disappear from the relative dynamics.
The energy rate therefore becomes
\begin{equation}
	\dot W=-D\Delta\omega^2-K_i z\Delta\omega.
	\label{eq:adaptive_energy_rate}
\end{equation}
The proportional term is dissipative for $D>0$.
Over an interval with $\Delta\omega>0$, both the power angle
and the integral state increase.
For $K_i>0$, the increase in $z$ relative to its value
at the start of the interval provides an additional
negative contribution to the energy rate.

%Changes in grid frequency are incorporated through the
%reference added in \eqref{eq:vsg_frequency_reconstruction}.
%This implementation does not explicitly differentiate the estimated frequency.

\subsection{Practical Considerations}

Next, we examine how errors introduced by grid impedance
uncertainty and filtering affect frequency estimation
and feedback control.

In a synchronous steady state, the positive-sequence voltage
and current rotate at the same frequency.
With constant line parameter errors, the reconstructed grid
voltage can therefore be expressed as
\begin{equation}
	\hat{\bar V}_g^{dq}=c\,\bar V_g^{dq},
	\label{eq:steady_voltage_reconstruction_error}
\end{equation}
where $\hat{\bar V}_g^{dq}$ denotes the reconstructed voltage
and $c$ is a constant complex factor representing magnitude
and phase errors.
For $c\neq0$, ideal differentiation gives
\begin{equation}
	\hat{\omega}_g
	=\omega_N+\operatorname{Im}
	\left(
	\frac{c\,\dot{\bar V}_g^{dq}}
	{c\,\bar V_g^{dq}}
	\right)
	=\omega_g.
	\label{eq:steady_frequency_invariance}
\end{equation}
Thus, constant line parameter errors do not bias the ideal
frequency estimate in steady state, provided the reconstructed
grid voltage remains nonzero.

During transients, reconstruction errors and filtering can
affect the estimated frequency.
Let $\varepsilon_\omega=\hat{\omega}_g-\omega_g$ denote the
frequency estimation error, so that
$e_\omega=\Delta\omega-\varepsilon_\omega$.
Since $\dot e_\omega=\Delta\dot{\omega}-\dot{\varepsilon}_\omega$,
the relative dynamics become
\begin{align}
	M\Delta\dot{\omega}
	&=P_m-P_e(\delta)-D\Delta\omega-K_i z \notag\\
	&\quad+D\varepsilon_\omega+M\dot{\varepsilon}_\omega,
	\label{eq:relative_swing_with_estimation_error}
\end{align}
with $\dot z=\Delta\omega-\varepsilon_\omega$.
The corresponding energy rate is
\begin{align}
	\dot W
	&=-D\Delta\omega^2-K_i z\Delta\omega \notag\\
	&\quad+\left(D\varepsilon_\omega
	+M\dot{\varepsilon}_\omega\right)\Delta\omega.
	\label{eq:energy_rate_with_estimation_error}
\end{align}
The above expression shows that the estimation error affects
the energy rate through its value $\varepsilon_\omega$
and its rate of change $\dot{\varepsilon}_\omega$.
For disturbances that do not lead to system collapse,
the grid frequency deviation and its rate of change remain bounded.
We further assume that the estimation error and its derivative
are bounded:
\begin{equation}
	|\varepsilon_\omega|\leq\varepsilon_{\omega,\max},
	\qquad
	|\dot{\varepsilon}_\omega|\leq\dot{\varepsilon}_{\omega,\max},
	\label{eq:estimation_error_bounds}
\end{equation}
where the right-hand sides are constant bounds on the
error magnitude and its derivative magnitude, respectively.
Then
\begin{align}
	&-D\Delta\omega^2
	+\left(D\varepsilon_\omega
	+M\dot{\varepsilon}_\omega\right)\Delta\omega \notag\\
	&\quad\leq-D\Delta\omega^2 +\left(D\varepsilon_{\omega,\max}
	+M\dot{\varepsilon}_{\omega,\max}\right)|\Delta\omega|.
	\label{eq:estimation_error_damping_bound}
\end{align}
For $D>0$, this combined contribution is negative when
\begin{equation}
	|\Delta\omega|>
	\varepsilon_{\omega,\max}
	+\frac{M}{D}\dot{\varepsilon}_{\omega,\max}.
	\label{eq:estimation_error_damping_condition}
\end{equation}
Increasing $D$ lowers this threshold.

The estimation error affects the integral state through
$\dot z=e_\omega=\Delta\omega-\varepsilon_\omega$.
If $\Delta\omega>0$ and
$|\varepsilon_\omega|<\Delta\omega$
hold throughout an interval, then $\dot z>0$.
For $K_i>0$, the resulting increase in $z$
above its value at the start of the interval
provides an additional negative contribution to the energy rate.

The proposed feedback therefore provides a degree of robustness
to frequency estimation errors.

\section{Controller Hardware in the Loop Test}
\label{sec:chil}

\begin{figure}[!t]
	\centering
	\includegraphics[width=2.6in]{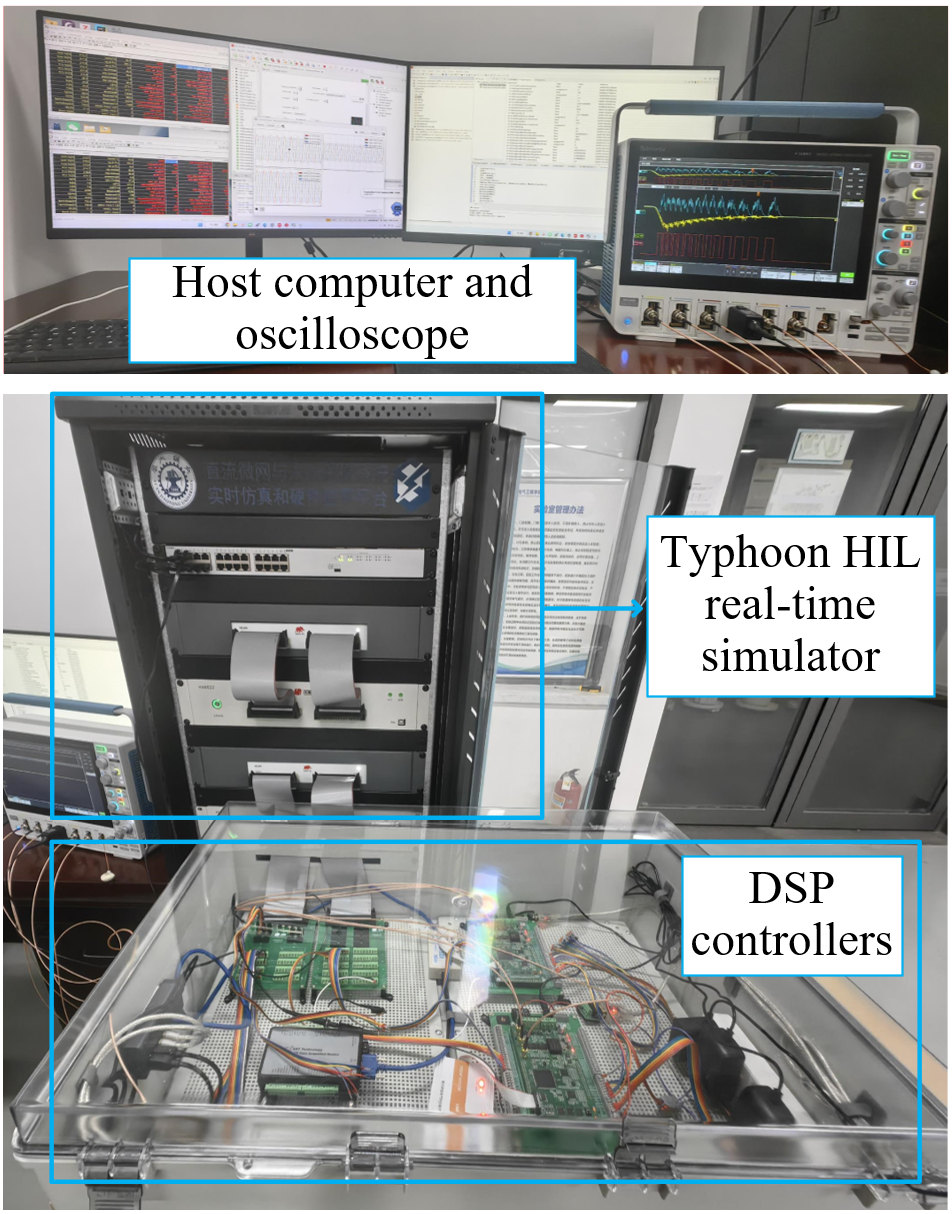}
	\caption{Controller hardware-in-the-loop experimental setup.}
	\label{CHIL}
\end{figure}

Controller hardware in the loop (CHIL) tests are conducted
to validate the proposed method under grid frequency disturbances.
The CHIL platform is shown in Fig.~\ref{CHIL}.
The power circuit is simulated on a Typhoon HIL404,
with two converters connected in parallel at the PCC.
The converter controllers are implemented on
TMS320F28377S DSPs.
The default grid and controller parameters are listed
in Table~\ref{tab:chil_parameters} in the Appendix.

All tests include fifth- and seventh-order voltage harmonics
of $5\%$ and $3\%$ of the fundamental voltage, respectively.
The default phase A, B, and C rms voltages are
$1.00$, $0.95$, and $1.05~\mathrm{p.u.}$, respectively. Cases C1 to C11 use one VSG and one GFL converter.
Case C12 uses two VSGs with different control parameters.
The tests cover frequency dips, combined frequency and
voltage disturbances, and changes in grid conditions,
power output, and controller settings.

The twelve cases are summarized in Table~\ref{tab:chil_cases}.
Each case compares two frequency references under the
same operating conditions and disturbance.
Panel (a) shows the estimation performance and system
response with $\omega_{\mathrm{ref}}=\omega_N$.
Panel (b) shows the response with
$\omega_{\mathrm{ref}}=\hat{\omega}_g$.
The grid frequency falls from $50$ to $48.3~\mathrm{Hz}$,
with a maximum decline rate of approximately
$1.8~\mathrm{Hz/s}$.
Cases C9 and C10 also include a voltage sag.

\begin{figure}[!t]
	\centering
	\includegraphics[width=3.3in]{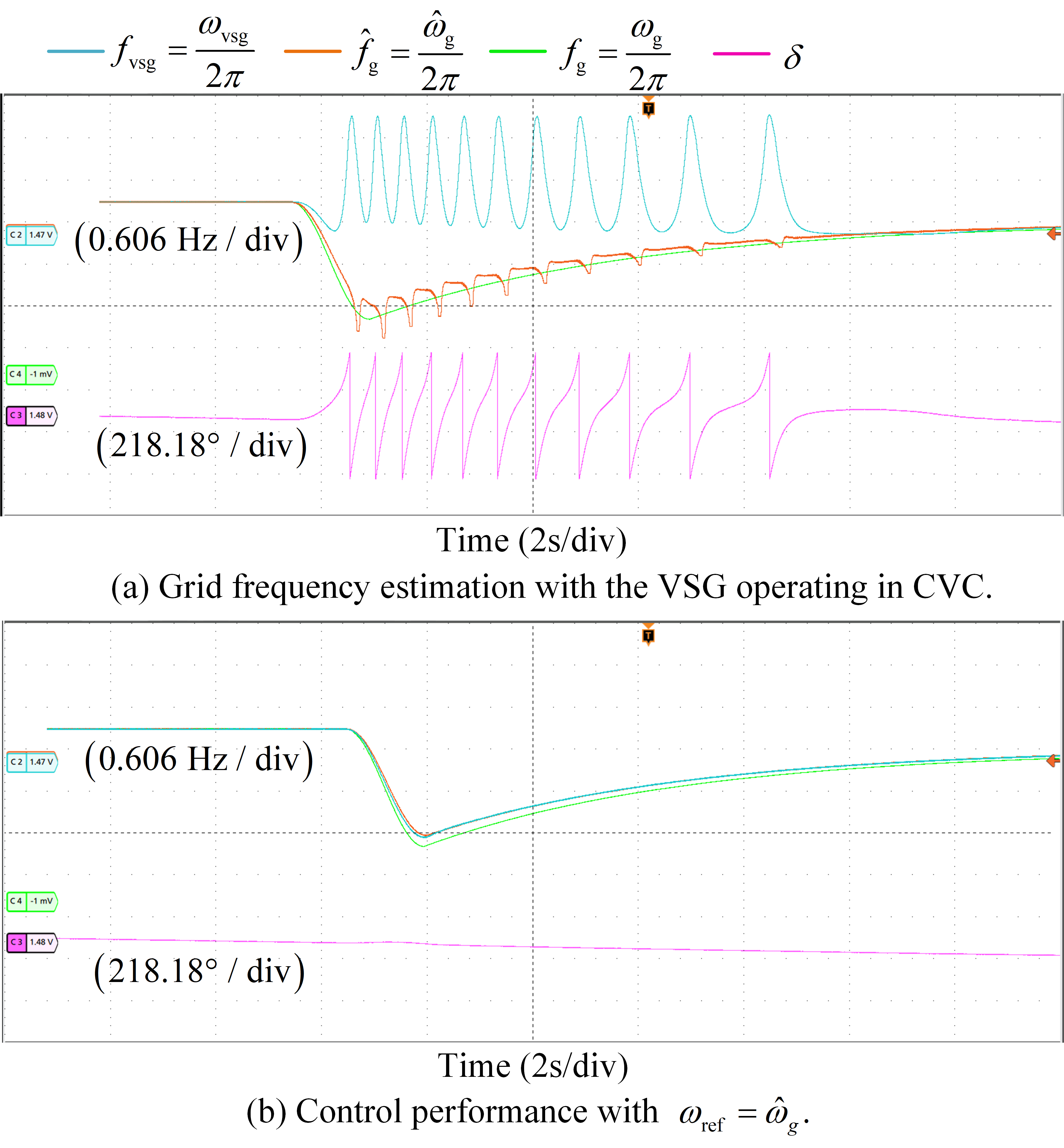}
	\caption{CHIL results for Case C1 with the VSG operating in CVC.}
	\label{C1}
\end{figure}

In Case C1, the VSG and GFL converter deliver
$5$ and $4.67~\mathrm{kW}$, respectively.
The VSG current limit is set to $3I_N$ so that it remains
in CVC during the disturbance.
As shown in Fig.~\ref{C1}(a), the VSG frequency
$f_{\mathrm{vsg}}$ does not follow the actual grid frequency
decline and exhibits large oscillations.
The accumulated frequency difference leads to loss of
synchronism.
The estimated frequency $\hat f_g$ follows the grid frequency
dip and recovery, with transient deviations during the
VSG oscillations.
With $\omega_{\mathrm{ref}}=\hat{\omega}_g$,
the VSG follows the grid frequency and its relative angle
remains bounded, as shown in Fig.~\ref{C1}(b).

\begin{figure}[!t]
	\centering
	\includegraphics[width=3.3in]{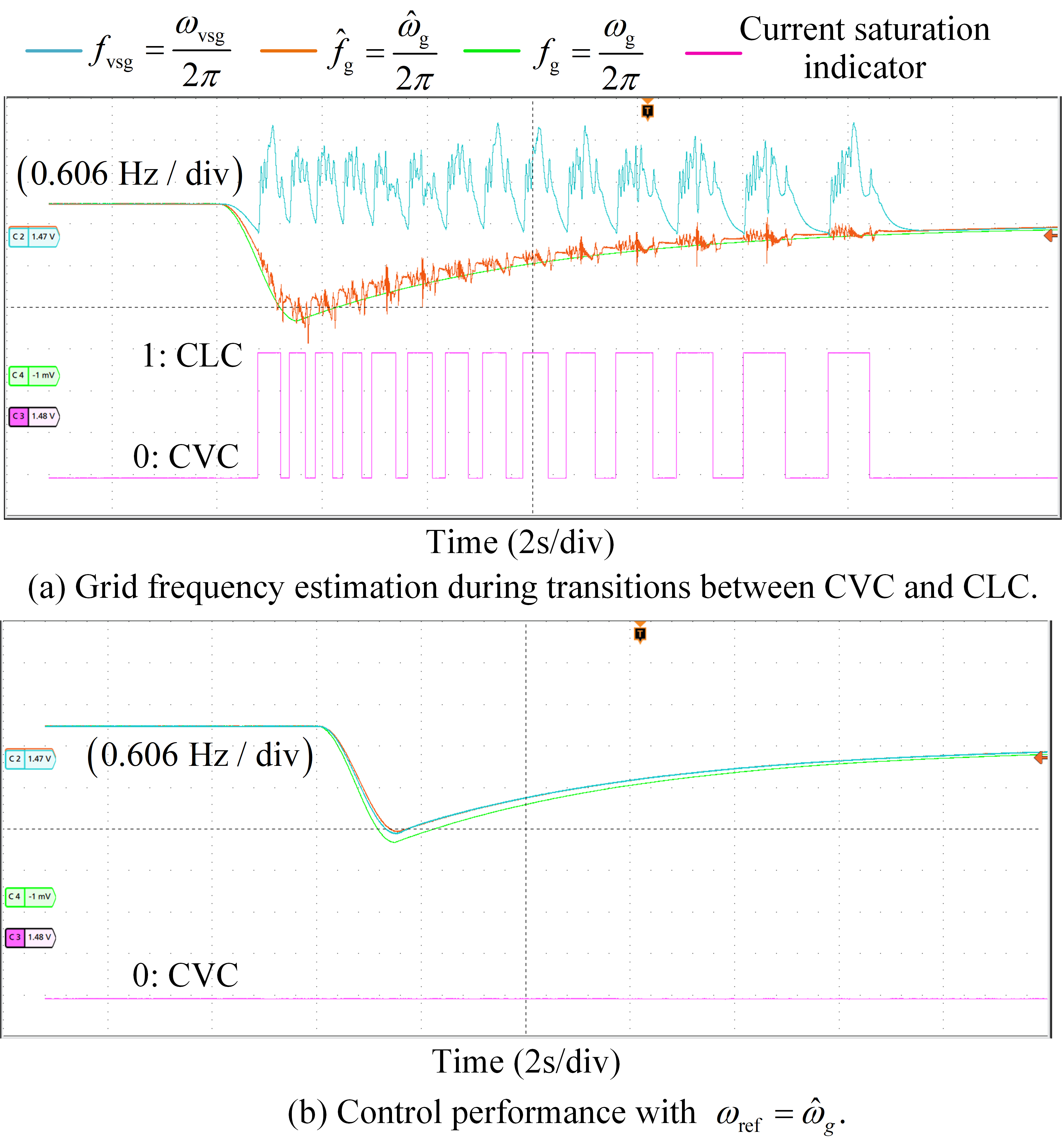}
	\caption{CHIL results for Case C2 with transitions between CVC and CLC.}
	\label{C2}
\end{figure}

Case C2 reduces the current limit to $1.5I_N$.
With the nominal frequency reference, the VSG repeatedly
switches between CVC and CLC, as indicated in
Fig.~\ref{C2}(a).
The estimate still captures the grid frequency trajectory,
although ripple increases during the mode transitions.
With the estimated frequency reference, the large frequency
oscillations are suppressed.
The VSG remains in CVC throughout the disturbance
in Fig.~\ref{C2}(b).

\begin{figure}[!t]
	\centering
	\includegraphics[width=3.3in]{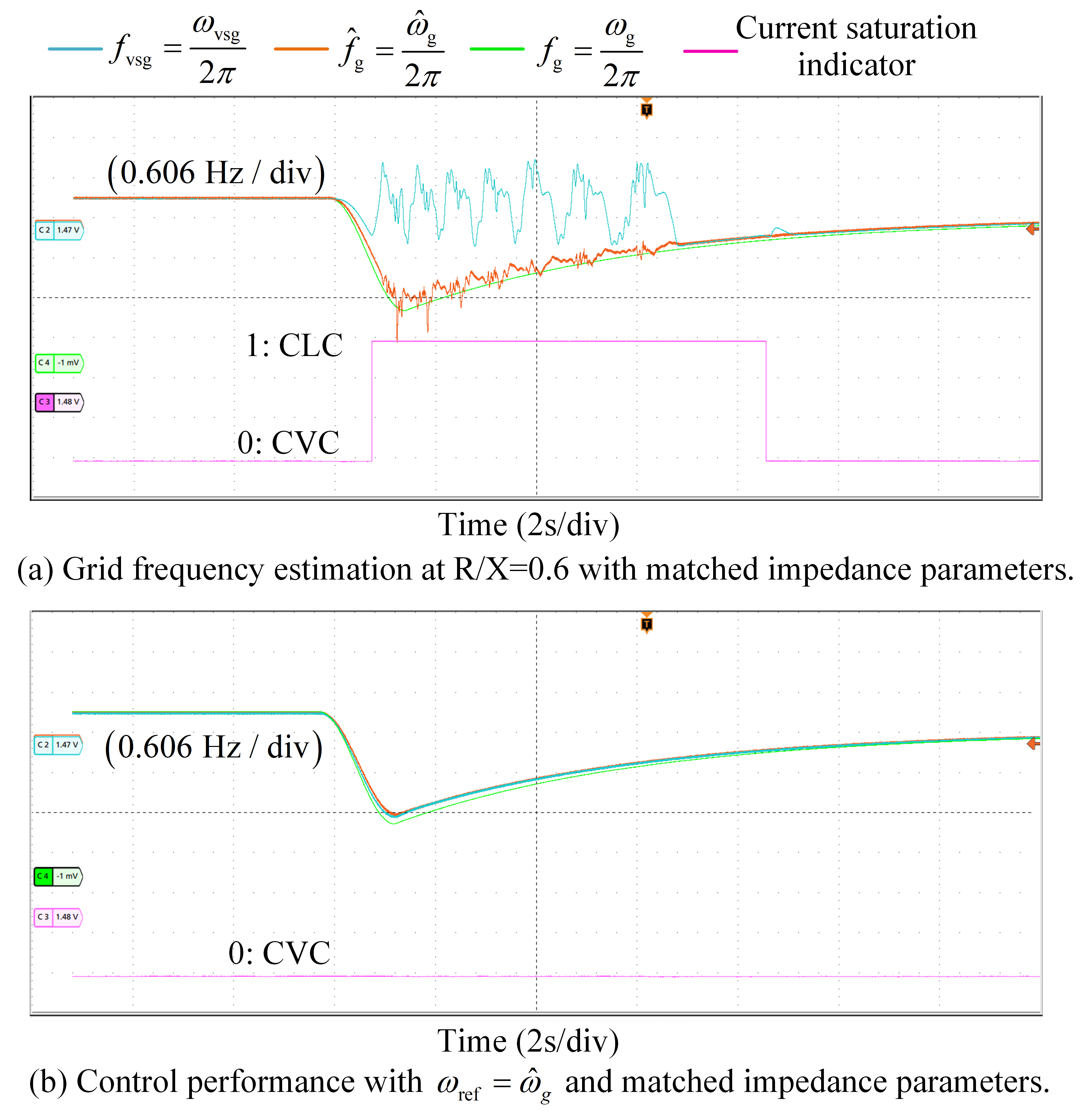}
	\caption{CHIL results for Case C3 at $R/X=0.6$
		with matched impedance parameters.}
	\label{C3}
\end{figure}

\begin{figure}[!t]
	\centering
	\includegraphics[width=3.3in]{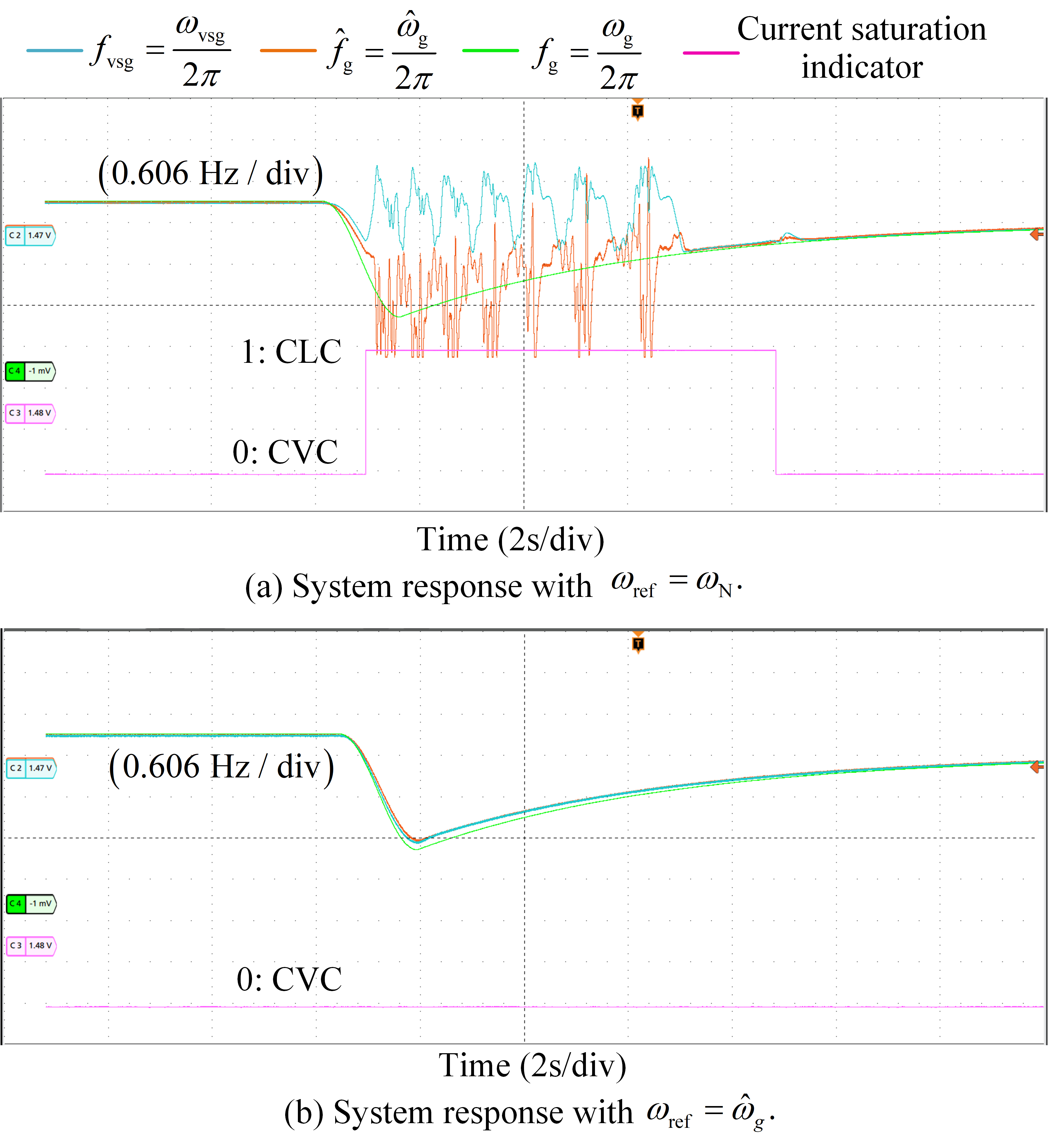}
	\caption{CHIL results for Case C4 at $R/X=0.6$
		with mismatched impedance parameters.}
	\label{C4}
\end{figure}

Cases C3 and C4 increase $R/X$ to $0.6$.
The estimator uses updated impedance parameters in C3
and retains the original parameters in C4.
With matched parameters, the estimate follows the grid
frequency during the extended CLC interval in
Fig.~\ref{C3}(a).
Parameter mismatch produces larger transient estimation
errors in Fig.~\ref{C4}(a).
Nevertheless, the estimated frequency feedback maintains
stable frequency tracking in both cases,
as shown in Figs.~\ref{C3}(b) and \ref{C4}(b).

\begin{figure}[!t]
	\centering
	\includegraphics[width=3.3in]{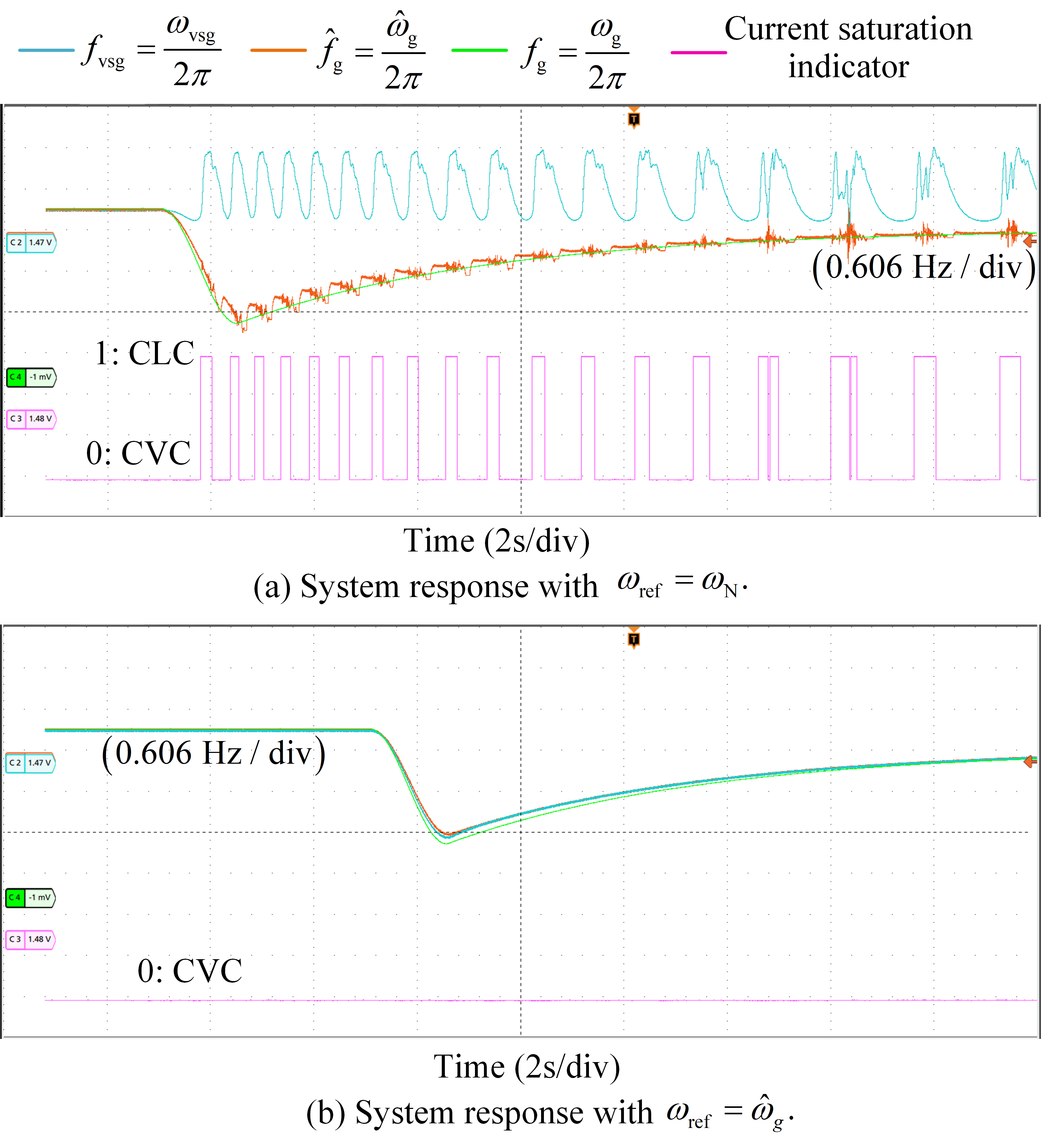}
	\caption{CHIL results for Case C5 at $\mathrm{SCR}=1.5$
		with matched impedance parameters.}
	\label{C5}
\end{figure}

\begin{figure}[!t]
	\centering
	\includegraphics[width=3.3in]{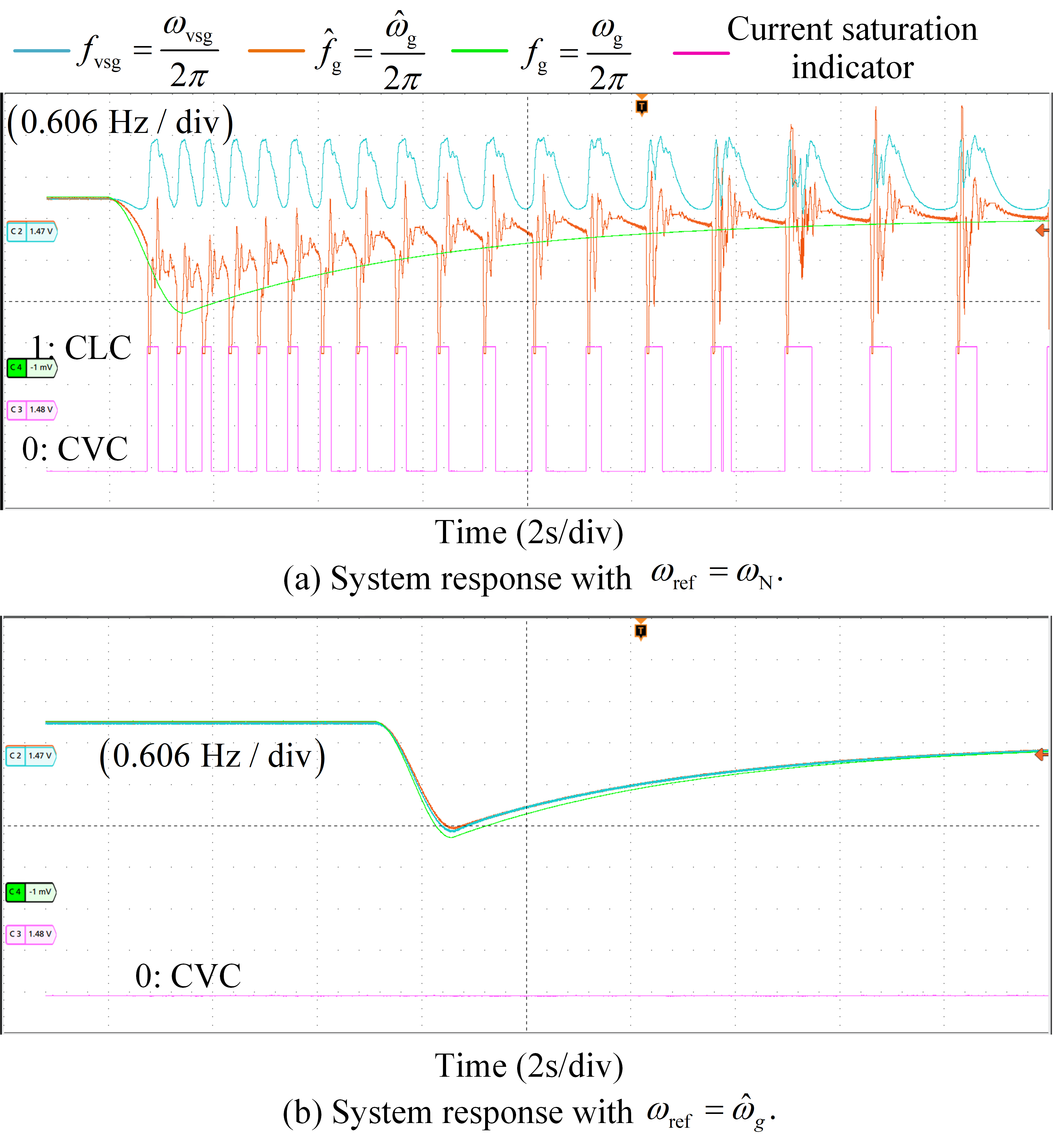}
	\caption{CHIL results for Case C6 at $\mathrm{SCR}=1.5$
		with mismatched impedance parameters.}
	\label{C6}
\end{figure}

Cases C5 and C6 reduce the SCR from $2.39$ to $1.5$
while keeping $R/X$ unchanged.
The impedance parameters used by the estimator are updated
in C5 and left unchanged in C6.
With the nominal reference, the VSG loses synchronism in both cases.
Its frequency oscillates, and it repeatedly switches between CVC and CLC.
The estimation errors are larger with mismatched parameters,
as shown in Fig.~\ref{C6}(a).
With the estimated frequency reference, the VSG follows
the grid frequency without entering CLC in either case,
as shown in Figs.~\ref{C5}(b) and \ref{C6}(b).
These results demonstrate effective control under weaker
grid conditions and impedance uncertainty.

\begin{figure}[!t]
	\centering
	\includegraphics[width=3.3in]{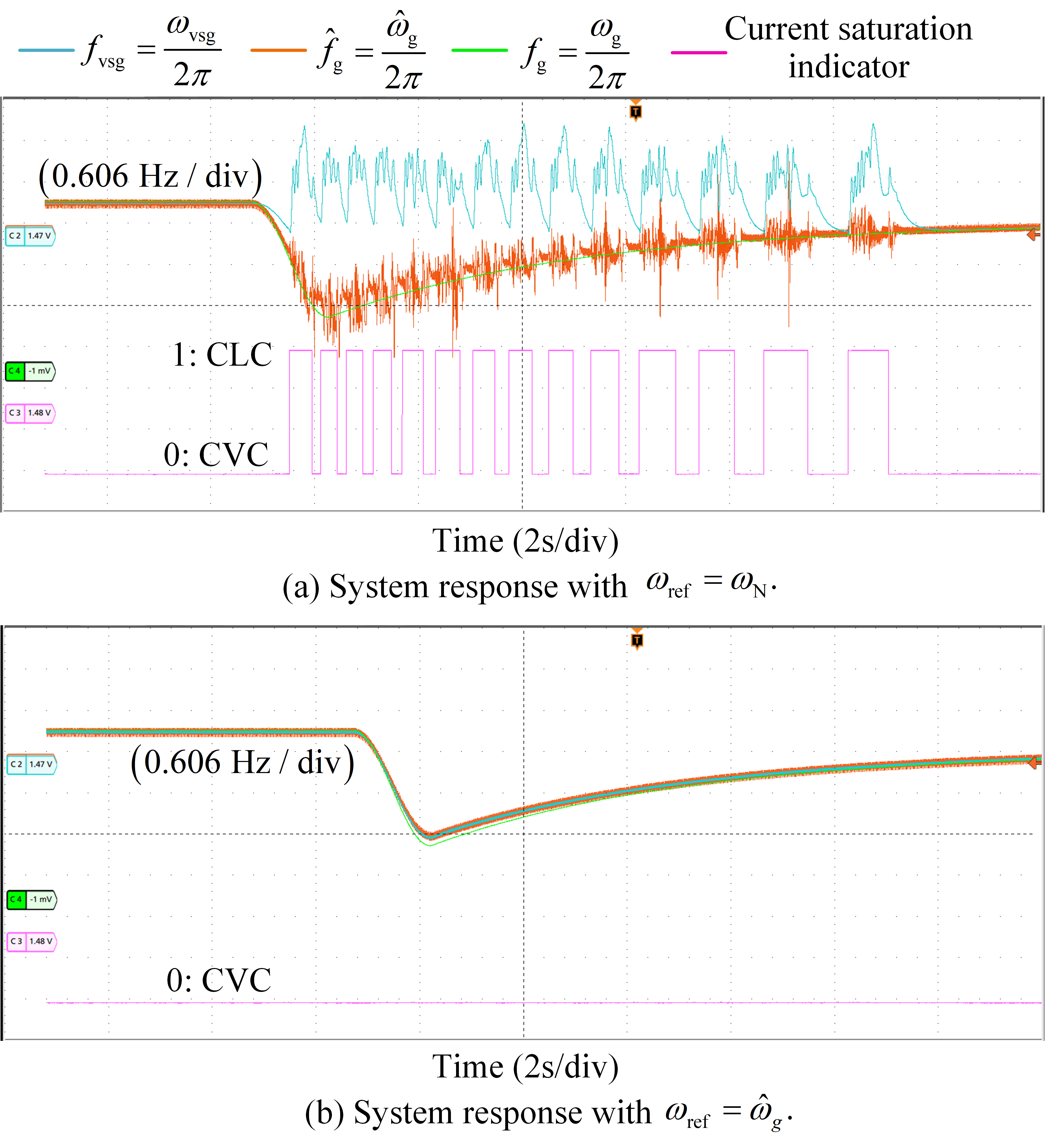}
	\caption{CHIL results for Case C7 with an estimated frequency
		filter cutoff of $20~\mathrm{Hz}$.}
	\label{C7}
\end{figure}

\begin{figure}[!t]
	\centering
	\includegraphics[width=3.3in]{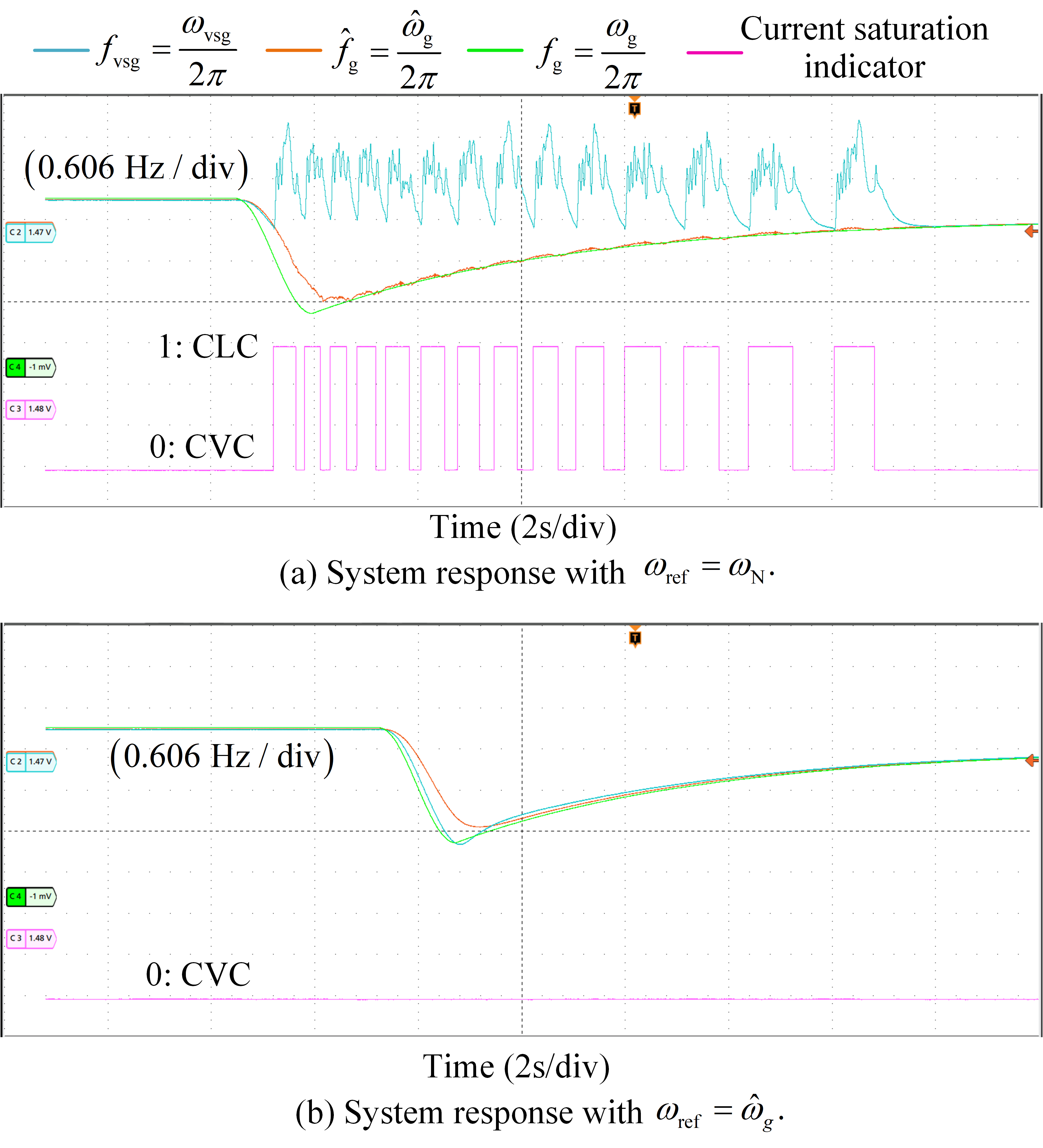}
	\caption{CHIL results for Case C8 with an estimated frequency
		filter cutoff of $0.5~\mathrm{Hz}$.}
	\label{C8}
\end{figure}

Cases C7 and C8 change the cutoff frequency of the final
frequency estimation filter from $5~\mathrm{Hz}$ to
$20$ and $0.5~\mathrm{Hz}$, respectively.
The higher cutoff allows more ripple in the estimate,
as shown in Fig.~\ref{C7}(a).
The lower cutoff produces a smoother estimate but introduces
a visible delay during the frequency decline
in Fig.~\ref{C8}(a).
Despite these differences, the estimated frequency feedback
suppresses the large VSG oscillations in both cases.
The responses in Figs.~\ref{C7}(b) and \ref{C8}(b)
remain stable.

\begin{figure}[!t]
	\centering
	\includegraphics[width=3.3in]{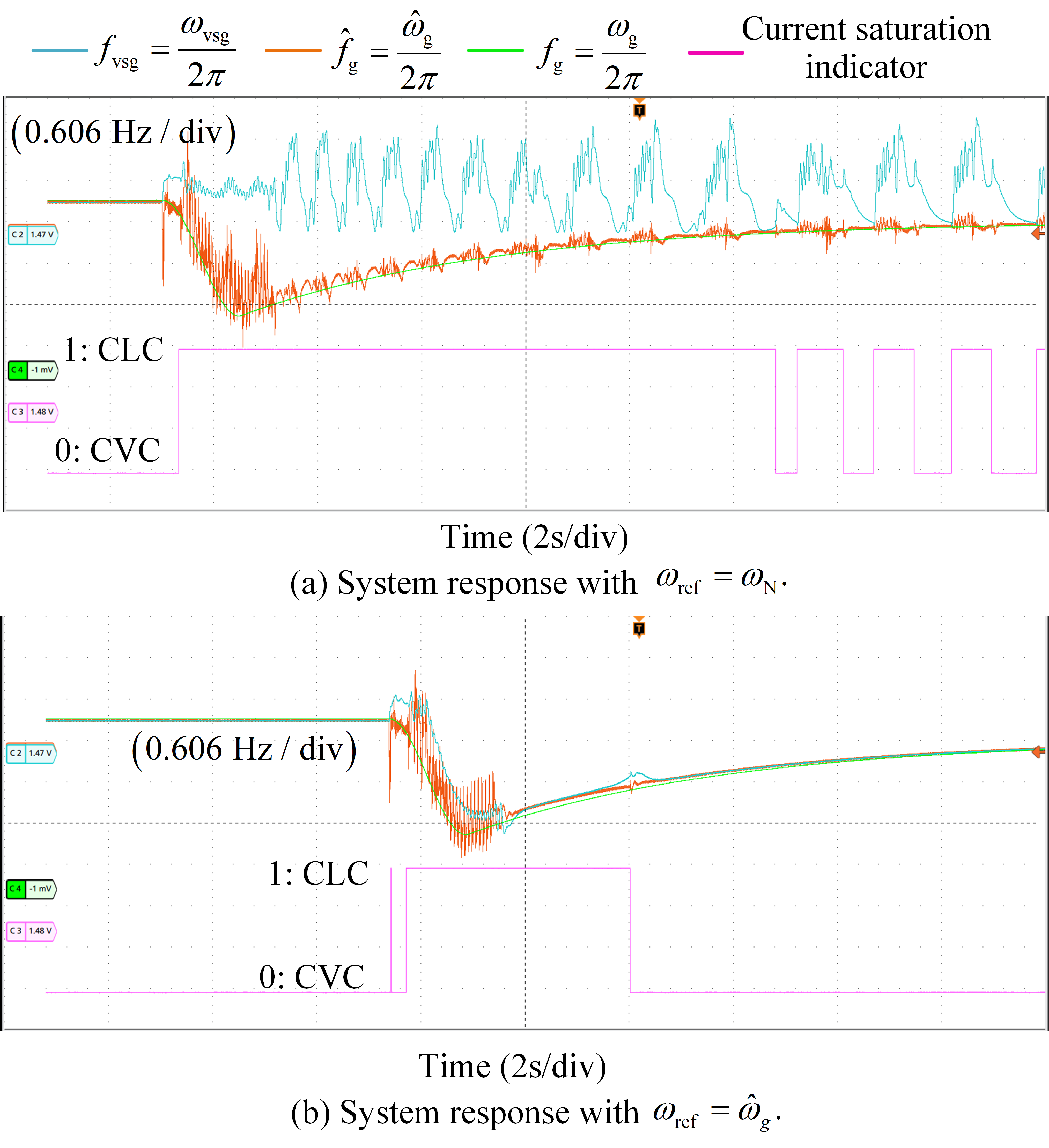}
	\caption{CHIL results for Case C9 with
		increased $P_{\mathrm{VSG}}$
		under a frequency dip and a severe balanced voltage sag.}
	\label{C9}
\end{figure}

Case C9 increases the VSG output to $8~\mathrm{kW}$
and combines the frequency dip with a balanced voltage sag.
All three phase voltages fall to $0.1~\mathrm{p.u.}$
and remain at this level for $2~\mathrm{s}$.
With the nominal reference, the VSG loses synchronism,
as shown in Fig.~\ref{C9}(a).
With the estimated frequency reference, the voltage sag
still causes current saturation,
as shown in Fig.~\ref{C9}(b).
However, the VSG subsequently returns to CVC
and follows the recovering grid frequency.

\begin{figure}[!t]
	\centering
	\includegraphics[width=3.3in]{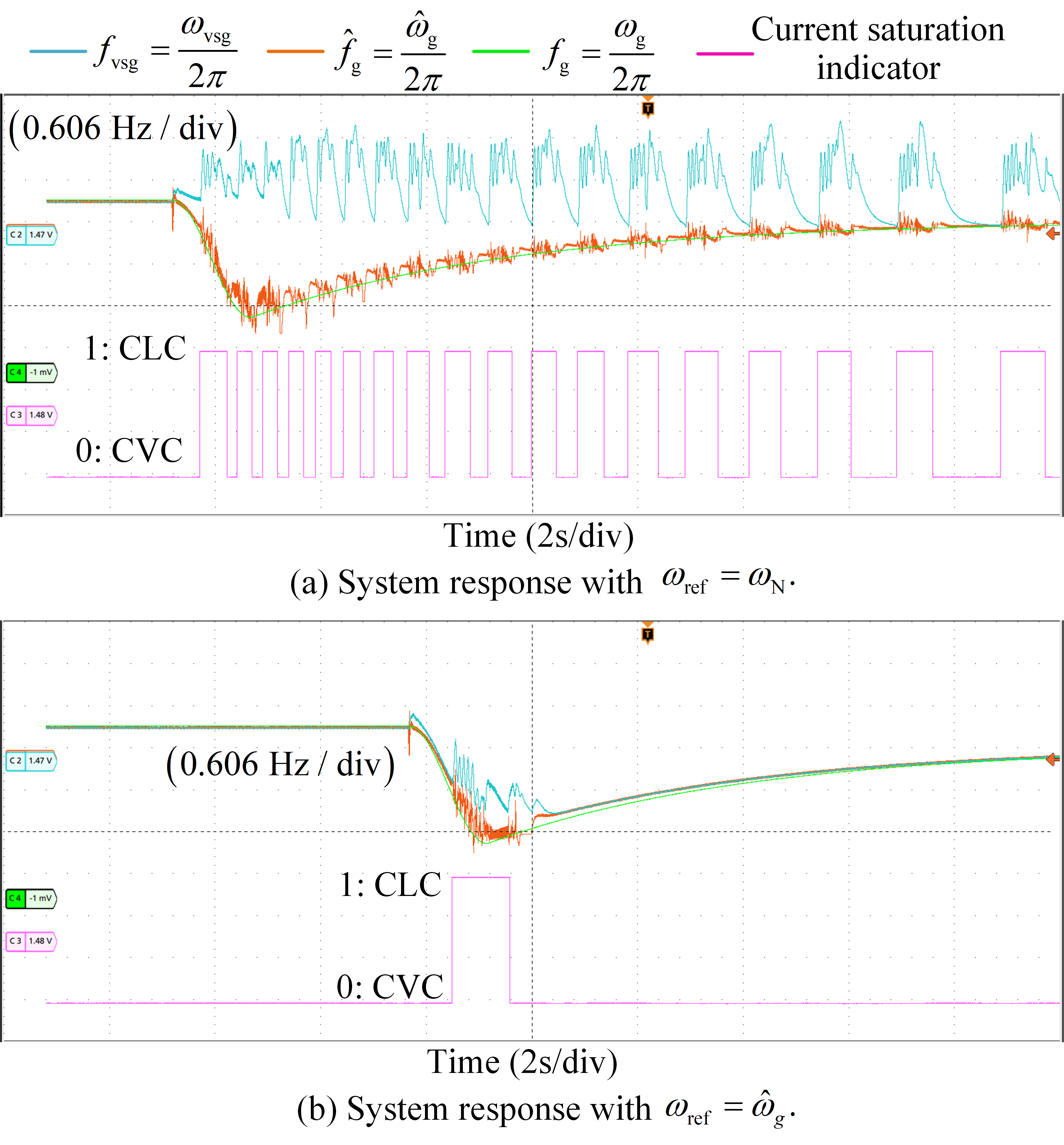}
	\caption{CHIL results for Case C10 with
		increased $P_{\mathrm{GFL}}$
		under a frequency dip and an unbalanced voltage sag.}
	\label{C10}
\end{figure}

In Case C10, the GFL output is increased to $7.47~\mathrm{kW}$.
The disturbance combines a frequency dip with an unbalanced
voltage sag.
The phase A, B, and C voltages fall to
$0.30$, $0.285$, and $0.315~\mathrm{p.u.}$, respectively,
and remain at these levels for $2~\mathrm{s}$.
With the nominal reference, the VSG loses synchronism
and experiences repeated current saturation,
as shown in Fig.~\ref{C10}(a).
With the estimated frequency reference, the VSG briefly
enters CLC and then returns to CVC.
The transient oscillations decay, and stable frequency
tracking is restored, as shown in Fig.~\ref{C10}(b).

\begin{figure}[!t]
	\centering
	\includegraphics[width=3.3in]{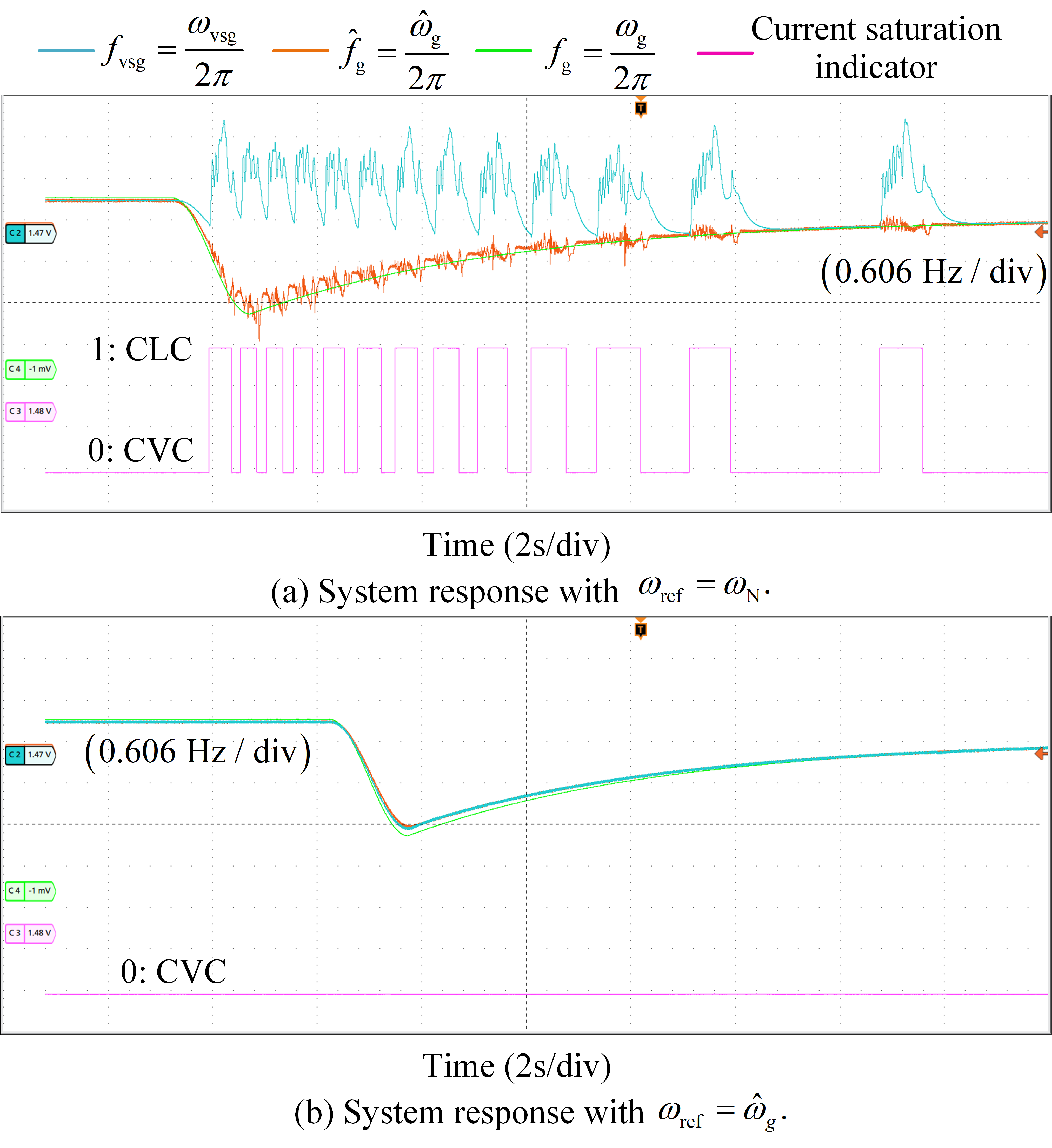}
	\caption{CHIL results for Case C11 with additional
		frequency PI feedback.}
	\label{C11}
\end{figure}

Case C11 adds frequency PI feedback with
$k_{p\omega}=50$ and $k_{i\omega}=500$.
With $\omega_{\mathrm{ref}}=\omega_N$,
the VSG exhibits large frequency oscillations
and repeated current saturation after losing synchronism,
as shown in Fig.~\ref{C11}(a).
Replacing the nominal reference with $\hat{\omega}_g$
maintains synchronism and keeps the VSG in CVC,
as shown in Fig.~\ref{C11}(b).
The proposed reference modification therefore remains
effective with additional integral feedback.

\begin{figure}[!t]
	\centering
	\includegraphics[width=3.3in]{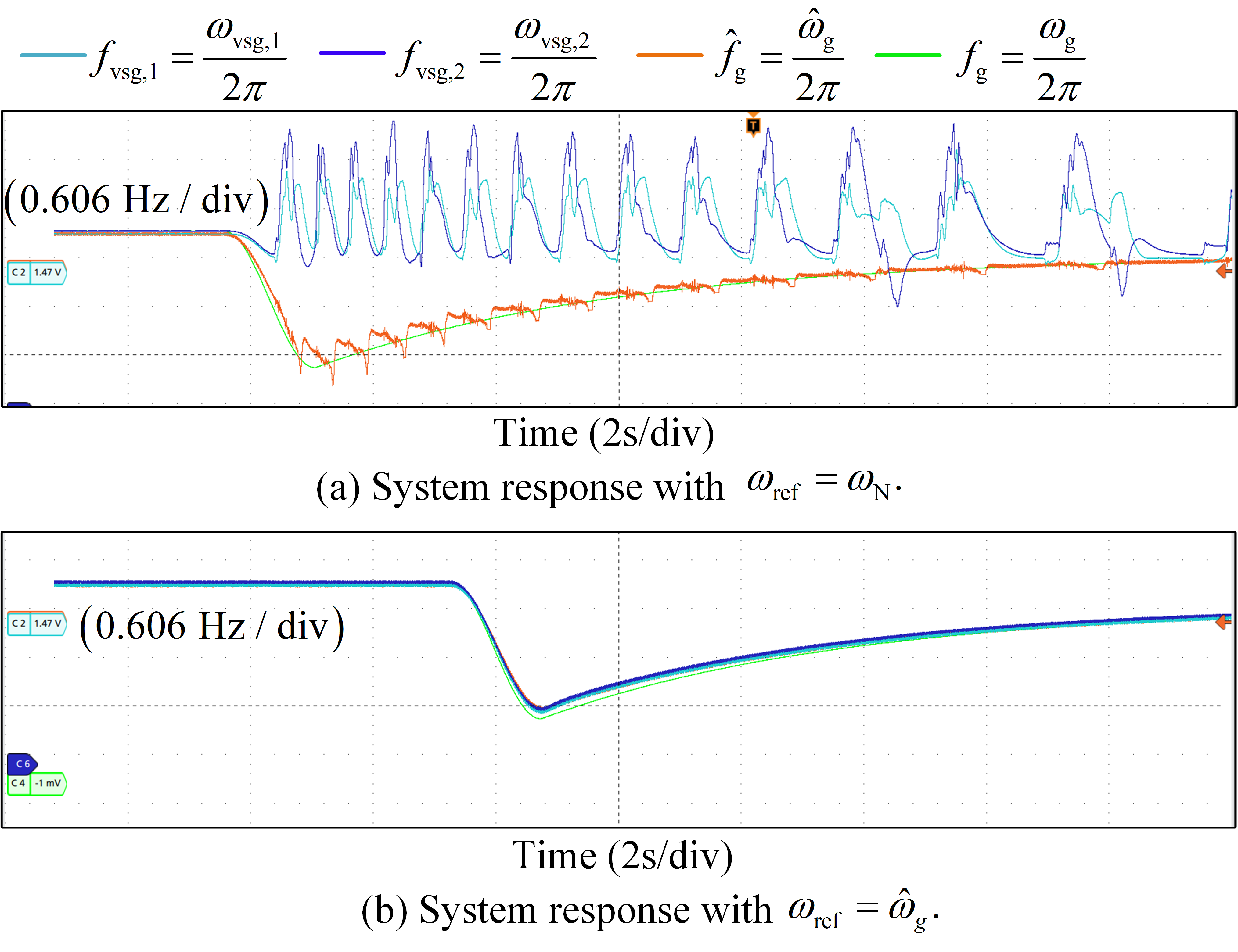}
	\caption{CHIL results for Case C12 with two parallel VSGs
		of different inertia and damping.}
	\label{C12}
\end{figure}

Case C12 replaces the GFL converter with a second VSG.
The two VSGs deliver $5$ and $3~\mathrm{kW}$
and have different inertia, damping, and current limits.
With the nominal reference, both VSGs lose synchronism
with the grid, as shown in Fig.~\ref{C12}(a).
The estimator still captures the grid frequency
dip and recovery.
With the estimated frequency reference, both VSGs
closely follow the grid frequency,
as shown in Fig.~\ref{C12}(b).
Stable operation is thus achieved with two parallel VSGs
despite their different parameters.

These tests verify grid frequency estimation in both
CVC and CLC.
Impedance mismatch and filter settings affect transient
estimation accuracy, but the proposed feedback maintains
or restores stable frequency tracking in all twelve cases.
The results demonstrate robustness to the tested grid
and controller parameter variations, voltage harmonics,
and voltage unbalance.
The method is also effective under combined frequency
and voltage disturbances and with multiple parallel VSGs.

\section{Electromechanical Transient Simulation of the IEEE 39-Bus System}

\begin{figure}[!t]
	\centering
	\includegraphics[width=3in]{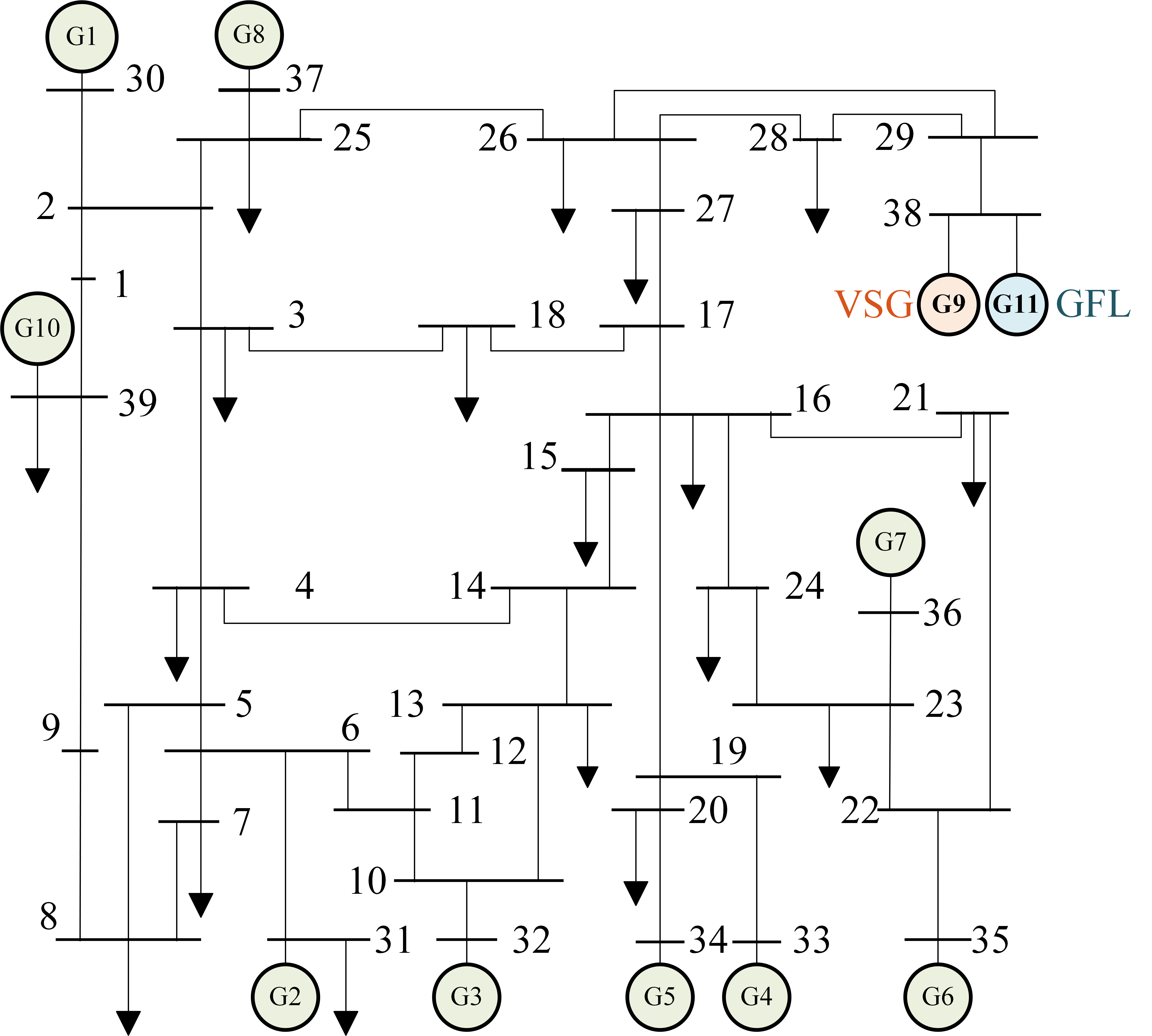}
	\caption{Topology of the modified IEEE 39-bus system.}
	\label{fig:ieee39}
\end{figure}

Electromechanical transient simulations are conducted in MATLAB
to evaluate the proposed grid frequency estimator and adaptive
frequency feedback in a multimachine system.
The modified IEEE 39-bus test system is shown in
Fig.~\ref{fig:ieee39}.
The network is represented by algebraic phasor equations
with constant-impedance loads.
The differential equations are solved using the modified
Euler method with a time step of $100~\mu\mathrm{s}$.

Gen~9, originally connected to bus~38, is replaced by a VSG.
A GFL converter, denoted Gen~11, is connected in parallel
at bus~38.
Their initial active power outputs in the prefault power flow
are $530~\mathrm{MW}$ and $400~\mathrm{MW}$, respectively.
The total active power generation and load are
$7270.35~\mathrm{MW}$ and $7112.17~\mathrm{MW}$, respectively.
The nominal frequency is $60~\mathrm{Hz}$.
The power and voltage bases are $100~\mathrm{MVA}$
and $345~\mathrm{kV}$.
The prefault Th\'evenin impedance of the external grid
seen from bus~38 is
$\bar{Z}=0.02474+j0.09341~\mathrm{p.u.}$.
The VSG has an inertia time constant of $6.9~\mathrm{s}$
and a damping coefficient of $1500~\mathrm{p.u.}$.
All rotor angles are measured relative to Gen~2 at bus~31.

At $t=10~\mathrm{s}$, the load is increased by
$4000~\mathrm{MW}$ and $667~\mathrm{Mvar}$.
Both increments are distributed among the load buses
in proportion to their initial active power demands.
The same disturbance and operating conditions are used
to evaluate frequency estimation and adaptive frequency feedback.

\begin{figure}[!t]
	\centering
	\includegraphics[width=\columnwidth]{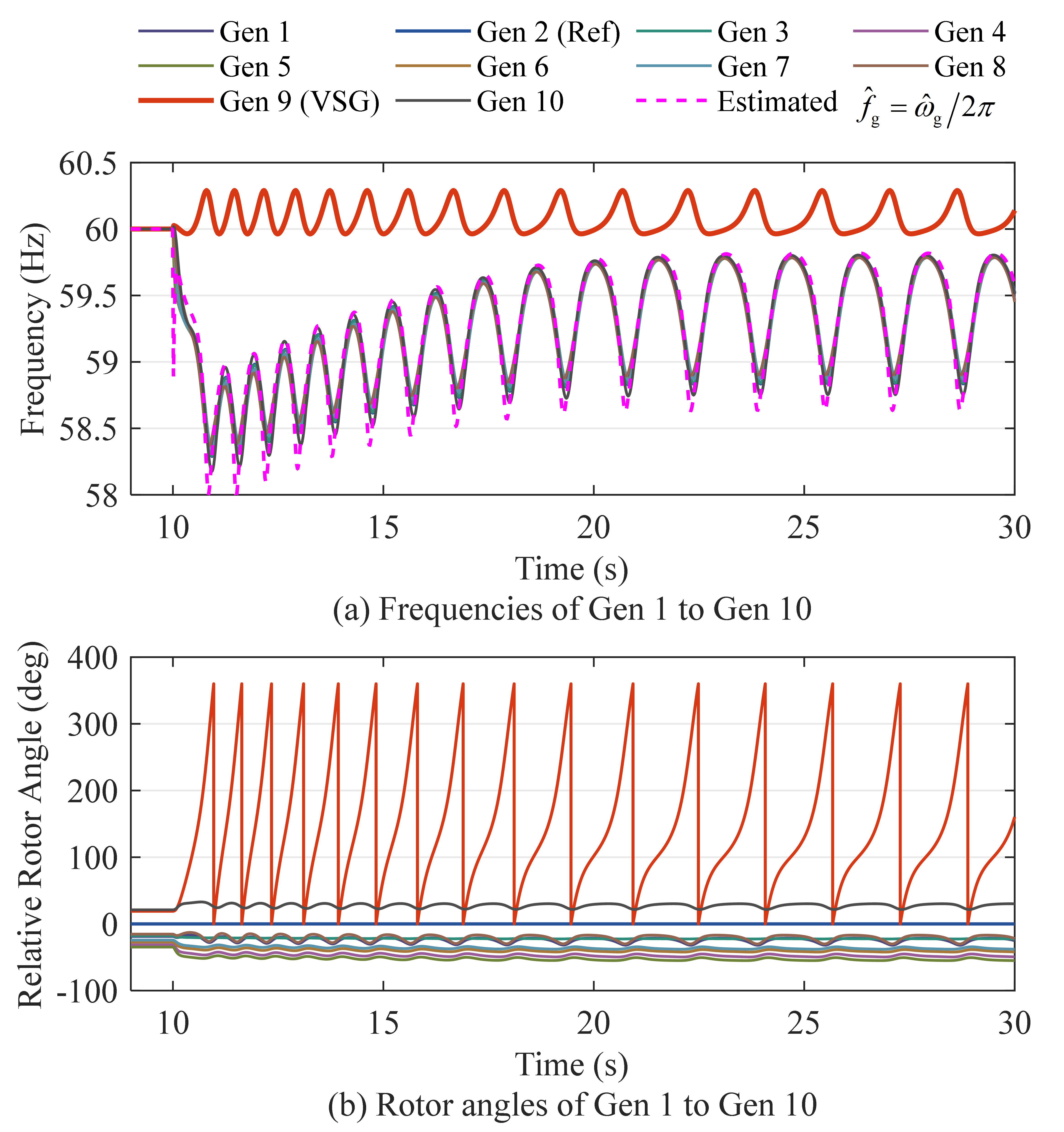}
	\caption{Grid frequency estimation performance.}
	\label{fig:ieee39_before_control}
\end{figure}

Fig.~\ref{fig:ieee39_before_control} shows the response
when the VSG frequency feedback uses the nominal reference
$\omega_{\mathrm{ref}}=\omega_N$.
Following the load increase, the synchronous generator
frequencies fall.
The large damping coefficient keeps the VSG frequency
near the nominal value.
A persistent frequency difference develops between the VSG
and the synchronous generators.
The VSG consequently loses synchronism, as shown in
Fig.~\ref{fig:ieee39_before_control}(b).
The proposed estimator follows the frequency decline
and subsequent oscillations of the synchronous generators,
with brief deviations during the transient.
It therefore captures the grid frequency response
even when the VSG loses synchronism.

\begin{figure}[!t]
	\centering
	\includegraphics[width=\columnwidth]{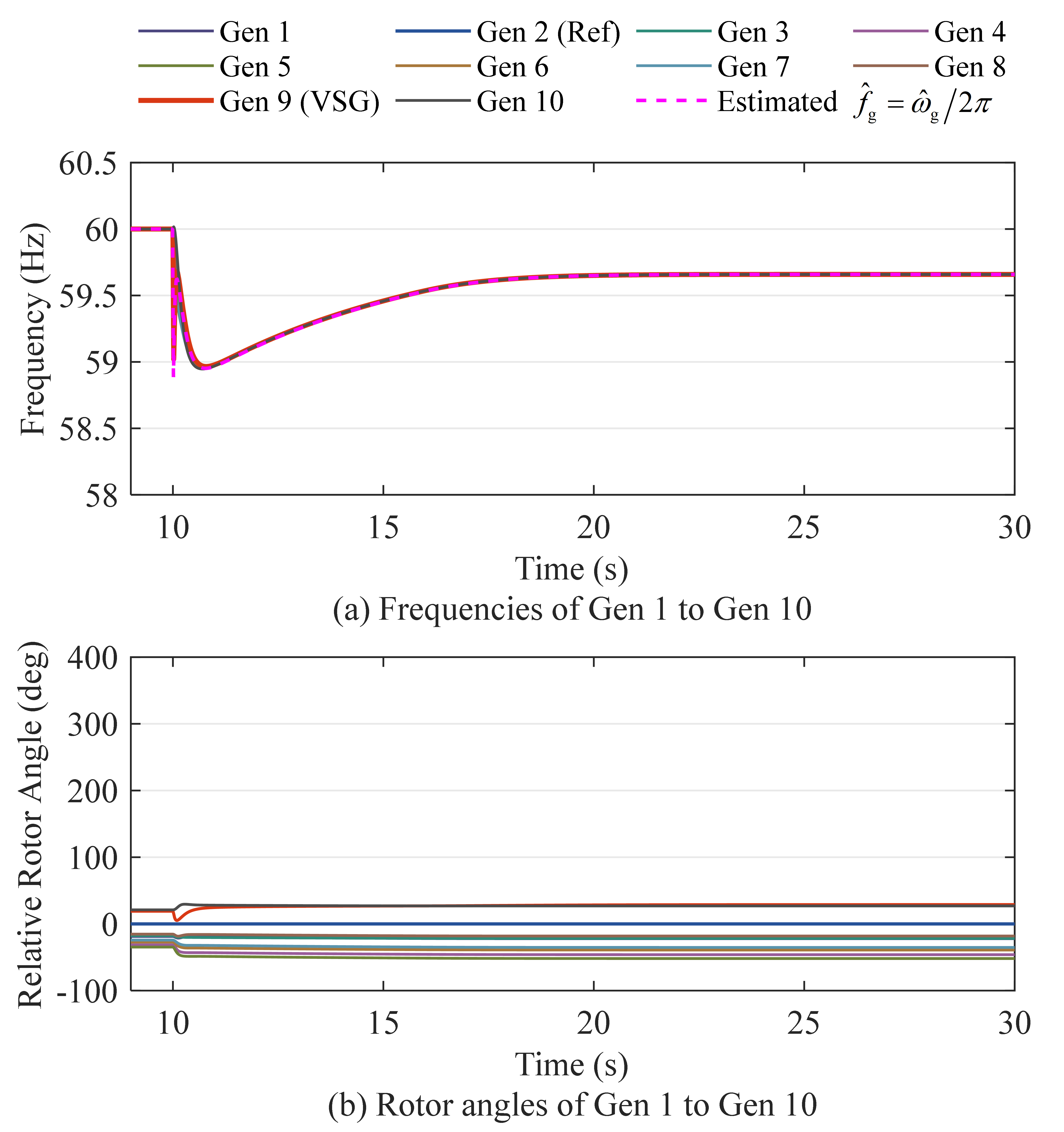}
	\caption{Transient response with the estimated grid frequency as the VSG frequency reference.}
	\label{fig:ieee39_after_control}
\end{figure}

The nominal reference $\omega_N$ in the VSG active power loop
is then replaced by the estimated grid frequency
$\hat{\omega}_g$, as defined in
\eqref{eq:adaptive_frequency_reference}.
The resulting response is shown in
Fig.~\ref{fig:ieee39_after_control}.
The VSG and synchronous generators converge to a common
frequency below $60~\mathrm{Hz}$.
Their relative rotor angles settle to constant values,
indicating that synchronism is maintained.
These results demonstrate the effectiveness of the proposed
estimator and adaptive frequency feedback in the
multimachine system.

\section{Conclusion}

This paper investigates a mechanism of VSG synchronization
instability during frequency dips in weak grids.
When damping feedback or emulated fast primary frequency
control uses the nominal frequency as its reference,
it can sustain a frequency difference between the VSG
and the grid.
The accumulated frequency difference can then lead
to power angle instability.
To address this issue, a grid frequency estimator
is developed based on complex frequency theory.
It uses only voltage and current measurements at the PCC,
without additional measurement devices or remote communication.
The estimated grid frequency replaces the nominal reference
in the VSG frequency feedback loop to improve transient stability.
CHIL tests validate the proposed method with a VSG
operating in parallel with a GFL converter
and with two parallel VSGs.
The tests verify its effectiveness in both CVC and CLC
and its robustness to grid and controller parameter
variations and impedance uncertainty.
Effective control is also demonstrated under voltage
harmonics and balanced or unbalanced voltage sags.
Simulations of the modified
IEEE 39-bus system further confirm the effectiveness
of adaptive frequency feedback
in a multimachine network.

This control particularly suited to enhancing the transient synchronization stability of renewable energy converters connected to the main grid through weak electrical connections. Future work will further investigate the coordination between VSG synchronization stability and inertial support.

\textcolor{black}{}
    %{\appendices
	%\section*{Appendices}

	%\subsection{System Parameters}
	%\begin{table}[ht]
		%\centering
		%\caption{System Parameters in Section \uppercase\expandafter{\romannumeral4}}
		%\label{tab:1}
		%\label{tab:system_parameters} % 设置标签
		%\begin{tabular}{lll}
			%\toprule
			%\textbf{Symbol} & \textbf{Item} & \textbf{Value} \\
			%\midrule
			%$U_{g}$ & Grid bus voltage & 1 p.u. \\
			%$i_{c}^d$ & Active current of GFLR & 0.002 p.u. \\
			%$L_{c},L_s,L_g$ & Inductance of branches & 0.2,0.2,0.6 p.u. \\
			%$R_{c}.R_s,R_g$ & Resistance of branches & 0.01,0.01,0.05 p.u. \\
			%$\omega_g$ & System angular speed & 314 rad/s \\
			%$k_p,k_I$ & Proportional and ntegral gain of PLL & 0.2,20 \\
			%$k_I$ & Integral gain of PLL & 20\\
			%\bottomrule
		%\end{tabular}
	%\end{table}
	%}
	\appendices
	
	\section{System Parameters}
	\label{app:parameters}
	% Place this table in the Appendix.

\begin{table}[!t]
	\centering
	\caption{Default Parameters of the CHIL Platform}
	\label{tab:chil_parameters}
	\footnotesize
	\renewcommand{\arraystretch}{1.0}
	\setlength{\tabcolsep}{3pt}
	
	\begin{tabular*}{\columnwidth}
		{@{\extracolsep{\fill}}>{\raggedright\arraybackslash}p{0.42\columnwidth}cc@{}}
		\toprule
		Parameter & Symbol & Value \\
		\midrule
		
		\multicolumn{3}{@{}c}{\textit{Power circuit}} \\
		Power base
		& $S_{\mathrm{base}}$ & $15~\mathrm{kVA}$ \\
		Nominal phase voltage (rms)
		& $V_N$ & $220~\mathrm{V}$ \\
		Nominal frequency
		& $f_N$ & $50~\mathrm{Hz}$ \\
		DC voltage
		& $V_{\mathrm{dc}}$ & $750~\mathrm{V}$ \\
		Filter inductance
		& $L_f$ & $1~\mathrm{mH}$ \\
		Filter resistance
		& $R_f$ & $0.01~\Omega$ \\
		Filter capacitance
		& $C_f$ & $20~\mu\mathrm{F}$ \\
		Capacitor damping resistance
		& $R_d$ & $1~\Omega$ \\
		Converter branch resistances
		& $R_{c1},R_{c2}$ & $0.1,\,0.1~\mathrm{m}\Omega$ \\
		Converter branch inductances
		& $L_{c1},L_{c2}$ & $0.1,\,0.5~\mathrm{mH}$ \\
		Grid resistance
		& $R$ & $0.01~\Omega$ \\
		Grid inductance
		& $L$ & $20~\mathrm{mH}$ \\
		5th voltage harmonic
		& $V_5/V_1$ & $5\%$ \\
		7th voltage harmonic
		& $V_7/V_1$ & $3\%$ \\
		Phase A voltage (unbalanced)
		& $V_a/V_N$ & $1.00~\mathrm{p.u.}$ \\
		Phase B voltage (unbalanced)
		& $V_b/V_N$ & $0.95~\mathrm{p.u.}$ \\
		Phase C voltage (unbalanced)
		& $V_c/V_N$ & $1.05~\mathrm{p.u.}$ \\
		
		\addlinespace[3pt]
		\multicolumn{3}{@{}c}{\textit{VSG control}} \\
		% M is obtained from the current DSP update:
		% M = J_code * omega_N * T_s, with J_code = 80.
		Equivalent inertia coefficient
		& $M_0$ & $1.9635~\mathrm{W\,s^2/rad}$ \\
		Base damping coefficient
		& $D_0$ & $4000~\mathrm{W\,s/rad}$ \\
		Voltage feedback gain
		& $K_v$ & $1500~\mathrm{var/V}$ \\
		Reactive power integral gain
		& $K_q$ & $0.01~\mathrm{V/(var\,s)}$ \\
		Virtual resistance
		& $R_v$ & $2.2~\Omega$ \\
		Virtual inductance
		& $L_v$ & $6~\mathrm{mH}$ \\
		%Current reference limit (peak)
		%& $I_{\max}$ & $100~\mathrm{A}$ \\
		Current PI gains
		& $k_{pi}^{\mathrm{VSG}},k_{ii}^{\mathrm{VSG}}$
		& $3~\mathrm{V/A},\,330~\mathrm{V/(A\,s)}$ \\
		%Frequency feedback PI gains
		%& $k_{p\omega},k_{i\omega}$ & $50,\,5$ \\
		%Power filter cutoff frequency
		%& $f_P$ & $10~\mathrm{Hz}$ \\
		
		\addlinespace[3pt]
		\multicolumn{3}{@{}c}{\textit{GFL control}} \\
		Positive-sequence current PI gains
		& $k_{pi}^{+},k_{ii}^{+}$ & $3~\mathrm{V/A},\,330~\mathrm{V/(A\,s)}$ \\
		Negative-sequence current PI gains
		& $k_{pi}^{-},k_{ii}^{-}$ & $2~\mathrm{V/A},\,51~\mathrm{V/(A\,s)}$ \\
		PLL PI gains
		& $k_{p\mathrm{PLL}},k_{i\mathrm{PLL}}$
		& \shortstack{$0.7~\mathrm{rad/(V\,s)}$,\\
		$12.8~\mathrm{rad/(V\,s^2)}$} \\
		
		\addlinespace[3pt]
		\multicolumn{3}{@{}c}{\textit{Frequency estimation}} \\
		%Assumed grid resistance
		%& $\hat R$ & $0.0101~\Omega$ \\
		%Assumed grid inductance
		%& $\hat L$ & $20.1~\mathrm{mH}$ \\
		%Sequence decoupling filter cutoff
		%& $f_{\mathrm{seq}}$ & $20~\mathrm{Hz}$ \\
		Reconstructed voltage filter cutoff
		& $f_v$ & $50~\mathrm{Hz}$ \\
		Derivative filter cutoff
		& $f_d$ & $80~\mathrm{Hz}$ \\
		Estimated frequency filter cutoff
		& $f_{\omega}$ & $5~\mathrm{Hz}$ \\
		
		\addlinespace[3pt]
		\multicolumn{3}{@{}c}{\textit{Digital implementation}} \\
		Real-time simulation step
		& $T_{\mathrm{sim}}$ & $1~\mu\mathrm{s}$ \\
		Controller sampling period
		& $T_s$ & $78.125~\mu\mathrm{s}$ \\
		Switching frequency
		& $f_{\mathrm{sw}}$ & $12.8~\mathrm{kHz}$ \\
		
		\bottomrule
	\end{tabular*}
	
\end{table}

\end{document}